\documentclass[11pt]{scrartcl}

\usepackage{graphicx}
\usepackage{epsf}
\usepackage[utf8]{inputenc}
\usepackage[T1]{fontenc}
\usepackage[english]{babel}
\usepackage{lmodern}
\usepackage{amsmath,amssymb}
\usepackage{icomma}
\usepackage{bbm}
\usepackage{url}

\begin{document}
\title{Credit Risk Meets Random Matrices: Coping with Non-Stationary Asset Correlations}
\author{Andreas M{\"u}hlbacher\footnote{andreas.muehlbacher@uni-due.de}~ and Thomas Guhr\\{\normalsize Faculty of Physics, University of Duisburg-Essen, Lotharstr. 1, 47048 Duisburg, Germany}}
\maketitle

\begin{abstract}
We review recent progress in modeling credit risk for
  correlated assets. We start from the Merton model which default events and 
  losses are derived from the asset values at maturity. To estimate
  the time development of the asset values, the stock prices are used
  whose correlations have a strong impact on the loss distribution,
  particularly on its tails. These correlations are
  non-stationary which also influences the tails. We account for the
  asset fluctuations by averaging over an ensemble
  of random matrices that models the truly existing set of measured
  correlation matrices. As a most welcome side effect, this
  approach drastically reduces the parameter dependence of the loss
  distribution, allowing us to obtain very explicit results which show
  quantitatively that the heavy tails prevail over diversification
  benefits even for small correlations. We calibrate our random matrix
  model with market data and show how it is capable of grasping
  different market situations.  Furthermore, we present numerical
  simulations for concurrent portfolio risks, \textit{i.e.},~for the
  joint probability densities of losses for two portfolios.  For the
  convenience of the reader, we give an introduction to the Wishart
  random matrix model.
\end{abstract}

\section{Introduction}
\label{sec1}

To assess the impact of credit risk on the systemic stability of the
financial markets and the economy as a whole is of considerable
importance as the subprime crisis of 2007-2009 and the events
following the collapse of Lehman Brothers drastically
demonstrated~\cite{Hull2009}. Better credit risk estimation is
urgently called for.  A variety of different approaches exists,
see~\cite{Bielecki2004,Bluhm2003,Crouhy2000,Duffie1999,Ibragimov2007,Lando2008,
  McNeil2005,Heitfield2006,Glasserman2006,Mainik2013} for an overview.
Most of them fall into the
reduced-form~\cite{Duffie1999,chava,schonbucher2003credit} or
structural-approach class~\cite{merton74,elizalde}, a comprehensive
review is given in \cite{giesecke}. The problem to be addressed
becomes ultimately a statistical one, as loss distributions for
large portfolios of credit contracts have to be estimated. Typically,
they have a very heavy right tail which is due to either unusually
large single events such as the Enron bankruptcy or the simultaneous
occurrence of many small events as seen during the subprime
crisis. Reducing this tail would increase the stability of the
financial system as a whole. Unfortunately, the claim that
diversification can lower the risk of a portfolio is questionable or
even wrong, because often the correlations between the asset values
are ignored.  They are very important in a portfolio of credit
contracts, \textit{e.g.}, in the form of collateralized debt
obligations (CDOs). In detailed studies, it was shown that the
presence of even weak positive correlations diversification fails to
reduce the portfolio risk~\cite{Schonbucher2001,Glasserman2004} for
first passage models and for the Merton
model~\cite{Schafer2007,Koivusalo2012,Schmitt2014}.

Recently, progress has been made to analytically solve the Merton
model~\cite{merton74} in a most general setting of a correlated
market and even in the realistic case of fluctuating correlations
between the assets. The covariance and correlation matrix of asset
values changes in time~\cite{Zang2011,Song2011,Munnix2012,
  Sandoval2012} exhibiting an important example of the
non-stationarity which is always present in financial markets. The
approach we review 
here~\cite{Schmitt2013,Schmitt2014,Schmitt2015,Sicking2018} uses the
fact that the set of different correlation matrices \textit{measured}
in a smaller time window that slides through a longer dataset can be
modeled by an ensemble of \textit{random} correlation matrices.  The
asset values are found to be distributed according to a correlation
averaged multivariate
distribution~\cite{chetalova,Schmitt2013,Schmitt2014,Schmitt2015}. This
assumption is confirmed by detailed empirical studies.  Applied to the
Merton model, this ensemble approach drastically reduces, as a most
welcome side effect, the number of relevant parameters. We are left
with only two, an average correlation between asset values and a
measure for the strength of the fluctuations. The special case of zero
average correlation has been previously
considered~\cite{Munnix2014}. The limiting distribution for a
portfolio containing an infinite number of assets is also given,
providing a quantitative estimates for the limits of diversification
benefits. We also report results of Monte-Carlo simulations for the
general case of empirical correlation matrices that yield the Value at
Risk (VaR) and Expected Tail Loss (ETL).

Another important aspect are concurrent losses of different
portfolios.  Concurrent extreme losses might impact the solvencies of
major market participants, considerably enlarging the systemic risks.
From an investor's point of view, buying CDOs allows to hold a
``slice'' of each contract within a given
portfolio~\cite{duffieCDO,longstaffCDO,BenmelechCDO}. Such an
investor might be severely affected by significant concurrent credit
portfolio losses.  It is thus crucial to assess in which way and how
strongly the losses of different portfolios are coupled. In the
framework of the Merton model and the ensemble average, losses of two 
credit portfolios are studied which are composed of statistically dependent credit
contracts. Since correlation coefficients only give full information
in the case of Gaussian distributions, the statistical dependence of
these portfolio losses are investigated by means of copulas
\cite{nelsen}. The approach discussed here differs from the one given
in \cite{li}, as Monte-Carlo simulations of credit portfolio losses
with empirical input from S\&P 500 and Nikkei 225 are run and the
resulting empirical copulas are analyzed in detail. There are many other
aspects causing systemic risk such as fire sales spillover \cite{lillo}.

This review paper is organized as follows: In Sec.~\ref{sec2} we
introduce Random Matrix Theory for non-stationary asset correlations,
including a sketch of the Wishart model for readers not familiar with
random matrices. This approach is used in Sec.~\ref{sec3} to account
for fluctuating asset correlations in Credit Risk. In Sec.~\ref{sec4},
concurrent credit portfolio losses are discussed. Conclusions are
given in Sec.~\ref{sec5}.

\section{Random Matrix Theory for Non-Stationary Asset Correlations}
\label{sec2}

We sketch the salient features of the Wishart
model for correlation and covariance matrices in Sec.~\ref{sec21}. 
In Sec.~\ref{sec22}, we discuss a new
interpretation of the Wishart model as a model to describe the
non-stationarity of the correlations.

\subsection{Wishart Model for Correlation and Covariance Matrices}
\label{sec21}

Financial markets are highly correlated systems, and risk assessment
always requires knowledge of correlations or, more generally, mutual
dependencies. We begin with briefly summarizing some of the facts
needed in the sequel. To be specific, we consider stock prices and
returns, but correlations can be measured in the same way for all
observables that are given as time series. We are interested in, say,
$K$ companies with stock prices $S_k(t), \ k=1,\ldots,K$ as functions
of time $t$. The relative price changes over a fixed time interval
$\Delta t$, \textit{i.e.}, the returns
\begin{equation}
 r_k(t) = \frac{S_k(t + \Delta t)-S_k(t)}{S_k(t)}
\label{returns}
\end{equation}
are well-known to have distributions with heavy tails, the smaller
$\Delta t$, the heavier. The sampled Pearson correlation
coefficients are defined as
 \begin{eqnarray}
C_{kl}  &=& \left\langle  M_k(t) M_l(t) \right\rangle_T \nonumber\\
M_k(t) &=& \frac{r_k(t) - \langle r_k(t)\rangle_T}
                                              {\sigma_k}
\label{corr}
\end{eqnarray}
between the two companies $k$ and $l$ in the time window of length
$T$. The time series $M_k(t)$ are obtained from the $r_k(t)$ by
normalizing to zero mean and to unit variance, where the standard deviation $\sigma_k$ is
evaluated in the above mentioned time window. We define the $K\times
T$ rectangular data matrix $M$ whose $k$-th row is the time series
$M_k(t)$. The correlation matrix with entries $C_{kl}$ is the given by
\begin{equation}
C \ = \ \frac{1}{T} MM^\dagger \ ,
\label{corr2}
\end{equation}
where $^\dagger$ indicates the transpose. By definition, $C$ is real
symmetric and has non-negative eigenvalues.  We will also use the
covariance matrix $\Sigma = \sigma C \sigma$ where the diagonal matrix
$\sigma$ contains the standard deviations $\sigma_k, k=1,\ldots,K$. Setting
$A=\sigma M$, we may write   
\begin{equation}
\Sigma \ = \ \frac{1}{T} AA^\dagger 
\label{corr3}
\end{equation}
for the covariance matrix. We have to keep in mind that correlations 
or covariances only fully grasp the mutual dependencies if the multivariate 
distributions are Gaussian, which is not the case for returns if $\Delta t$ 
is too small. We come back to this point.

Correlation or covariance matrices can be
measured for arbitrary systems in which the observables are time
series. About ninety years ago,
Wishart~\cite{Wishart1928,muirhead2009aspects} put forward a random
matrix model to assess the statistical features of the correlation or
covariance matrices by comparing to a Gaussian null
hypothesis. Consider the $K$ values $A_k(t), \
k=1,\ldots,K$ at a fixed time $t$ which form $K$ component data vector $A(t)$. Now suppose
that we draw the entries of this vector from a multivariate Gaussian
distribution with some covariance matrix $\Sigma_0$, say, meaning that
\begin{equation}
\widetilde{w}(A(t)|\Sigma_0) \ = \ \frac{1}{\det^{1/2}(2\pi \Sigma_0)} 
                  \exp\left(-\frac{1}{2}
                       A^\dagger(t) \Sigma_0^{-1} A(t) \right) 
\label{wishart0}
\end{equation}
is the probability density function. We now make the important
assumptions that, first, the data vectors are statistically
independent for different times $t$ and, second, the
distribution~\eqref{wishart0} has exactly the same form for all times
$t=1,\ldots,T$ with the same covariance matrix $\Sigma_0$. Put
differently, we assume that the data are from a statistical viewpoint,
Markovian and stationary in time. The probability density function for
the entire model data matrix $A$ is then simply the product
\begin{eqnarray}
w(A|\Sigma_0) &=& \prod_{t=1}^T \widetilde{w}(A(t)|\Sigma_0) \nonumber\\
            &=& \frac{1}{\det^{T/2}(2\pi \Sigma_0)} 
             \exp\left(-\frac{1}{2}\text{tr} A^\dagger \Sigma_0^{-1} A \right) \ .
\label{wishart1}
\end{eqnarray}
This is the celebrated Wishart distribution for the data matrix $A$
which predicts the statistical features of random covariance matrices.
By construction, we find for the average of the model covariance matrix 
$AA^\dagger/T$
\begin{eqnarray}
\langle \frac{1}{T} AA^\dagger\rangle 
          \ = \ \int d[A] w(A|\Sigma_0) \frac{1}{T} AA^\dagger
          \ = \ \Sigma_0 \ ,
\label{wishart2}
\end{eqnarray}
where the angular brackets indicate the average over the Wishart
random matrix ensemble~\eqref{wishart1} and where $d[A]$ stands for
the flat measure, \textit{i.e.}, for the product of the differentials
of all independent variables. The Wishart ensemble is based on the
assumptions of statistical independence for different times,
stationarity and a multivariate Gaussian functional form. The
covariance matrix $\Sigma_0$ is the input for the mean value of the
Wishart ensemble about which the individual random covariance matrices
fluctuate in a Gaussian fashion. The strength of the fluctuations is
intimately connected with the length $T$ of the model time
series. Taking the formal limit $T\rightarrow\infty$ reduces the
fluctuations to zero, and all random covariance matrices are fixed to
$\Sigma_0$.  It is worth mentioning that the Wishart model for random
correlation matrices has the same form. If we replace $A$ with $M$ and
$\Sigma_0$ with $C_0$ we find the Wishart distribution that yields the
statistical properties of random correlation matrices.

The Wishart model serves as a bench mark and a standard tool in
statistical inference~\cite{muirhead2009aspects} by means of an
ergodicity argument: the statistical properties of \textit{individual}
covariance or correlation matrices may be estimated by an ensemble of
such matrices, provided their dimension $K$ is large. Admittedly, this
ergodicity argument does not necessarily imply that the probability
density functions are multivariate Gaussians.  Nevertheless, arguments
similar to those that lead to the Central Limit Theorem corroborate
the Gaussian assumption and empirically it was seen to be justified in
a huge variety of applications. A particularly interesting application
of the Wishart model for correlations in the simplified form with
$C_0=1_K$ was put forward by the Paris and Boston econophysics
groups~\cite{Laloux1999,Plerou1999a} who compared the eigenvalue
distributions (marginal eigenvalue probability density functions) of
empirical financial correlation matrices with the theoretical
prediction. They found good agreement in the bulk of the distributions
which indicates a disturbing amount of noise-dressing in the data due
to the relatively short lengths of the empirical time series with
considerable consequences for portfolio optimization
methods~\cite{BP2000,Plerou2002,Giada2002,GUKA2003,Pafka2004,Tum2005}.

\subsection{New Interpretation and Application of the Wishart Model}
\label{sec22}

\begin{figure}[htbp]
  \begin{center}
    \includegraphics[width=0.75\textwidth]{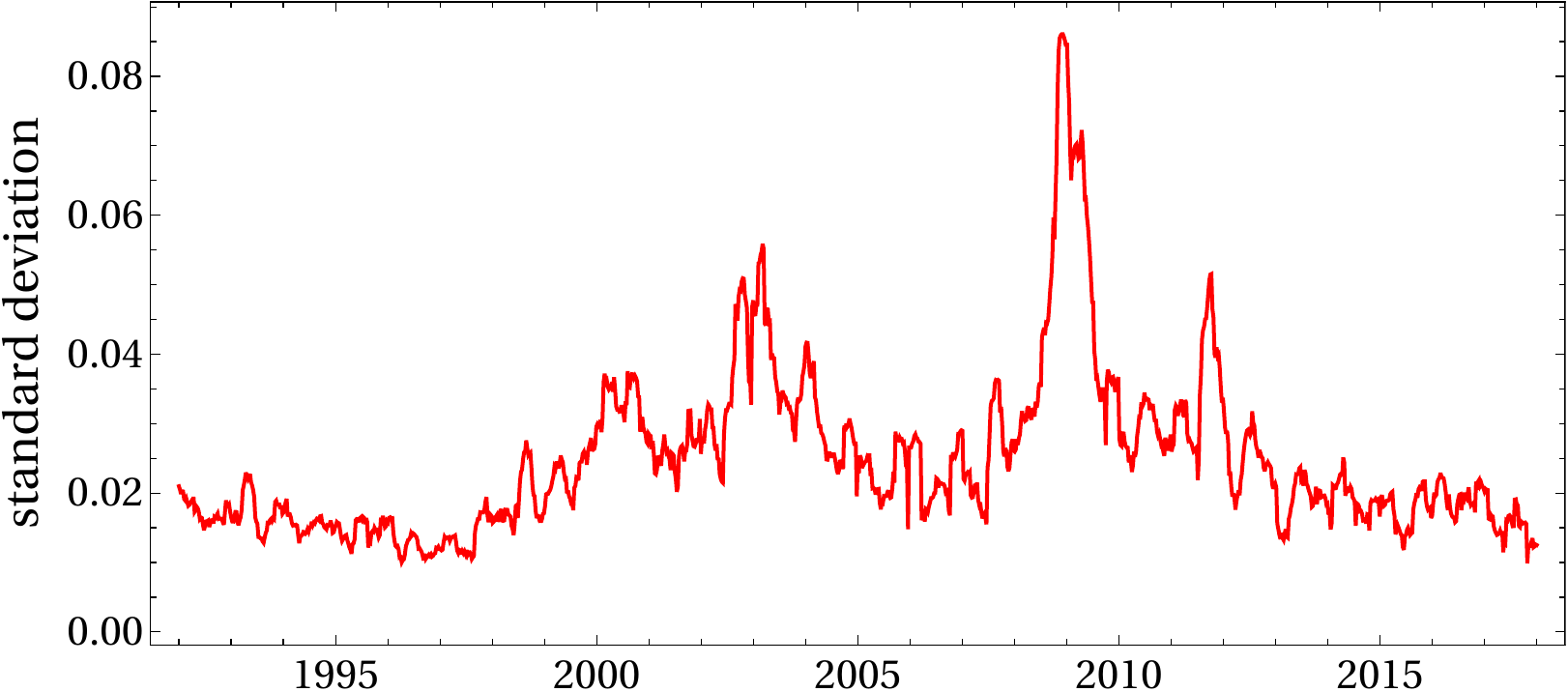}
  \end{center}
  \caption{Standard deviation time series for Goodyear from 1992 to 2018.
    The return interval is $\Delta t=1$ trading day and the time window has length
    $T= 60$ trading days.}
 \label{fig1}
\end{figure}

\begin{figure}[htbp]
  \begin{center}
    \includegraphics[width=0.3\textwidth]{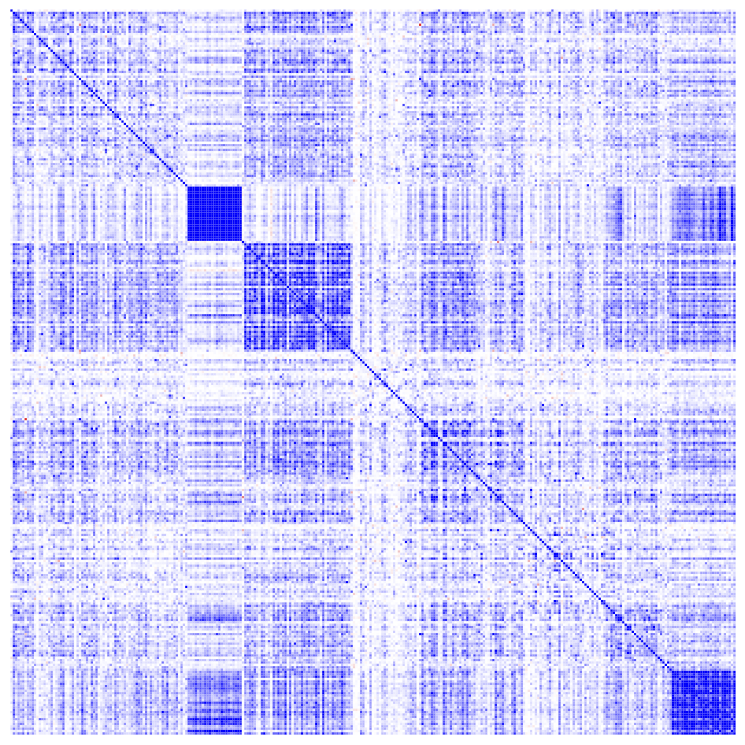}
    \includegraphics[width=0.3\textwidth]{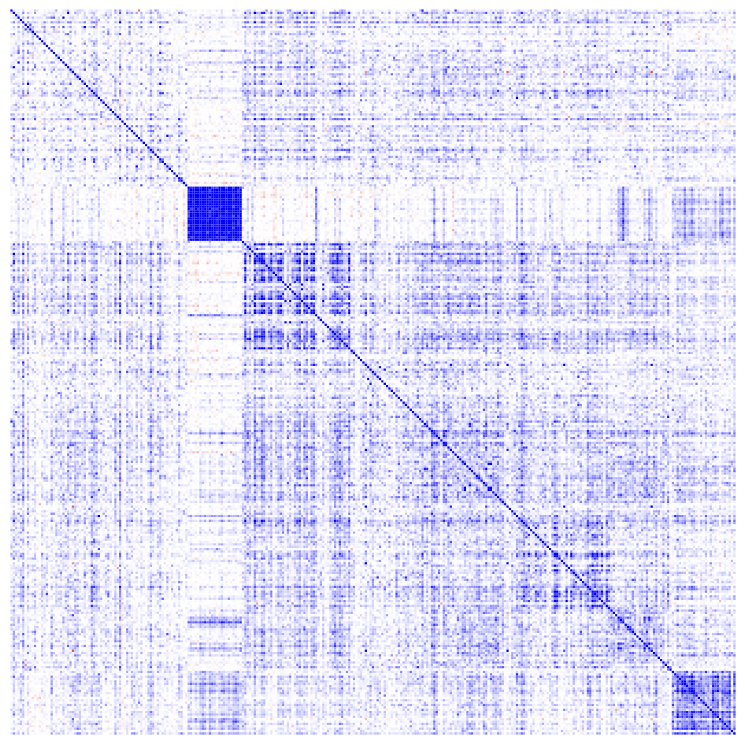}
  \end{center}
  \caption{Correlation matrices of $K=306$ companies for the fourth
    quarter of 2005 and the first quarter of 2006, the darker, the
    stronger the correlation. The companies are sorted according to
    industrial sectors. Taken from~\cite{Schmitt2013}.}
 \label{fig3}
\end{figure}

Financial markets are well-known to be non-stationary, \textit{i.e.},~the
assumption of stationarity is only meaningful on short time scales and is
bound to fail on longer ones. Non-stationary complex systems pose
fundamental
challenges~\cite{Gao1999,Hegger2000,Bernaola-Galvan2001,Rieke2002}
for empirical analysis and for mathematical
modeling~\cite{Zia2004,Zia2006}. An example from finance are the
strong fluctuations of the sample standard deviations $\sigma_k$,
measured in different time windows of the same length $T$
\cite{Black:1976fj,Schwert1989}, as shown in Fig.~\ref{fig1}.
Financial markets demonstrated their non-stationarity in a rather
drastic way during the recent years of crisis. Here, we focus on the
non-stationarity of the correlations.  Their fluctuations in time
occur, \textit{e.g.}, because the market expectations of the traders
change, the business relations between the companies change,
particularly in a state of crisis, and so on. To illustrate how
strongly the $K\times K$ correlation matrix $C$ as a whole changes in
time, we show it for subsequent time windows in Fig.~\ref{fig3}.
The dataset used here consists of
$K=306$ continuously traded companies in the S\&P 500 index between 1992
and 2012~\cite{yahoo}.  For later discussion, we emphasize that the
stripes in these correlation matrices indicate the structuring of the
market in industrial sectors, see, \textit{e.g.},~\cite{Munnix2012}.

Clearly, the non-stationary fluctuations of the correlations
influence all deduced economic observables, and it is quite plausible
that this effect will be strong for the statistics of rare, correlated
events. In the sequel we will show that the tails of the loss
distributions in credit risks will be particularly sensitive to the
non-stationarity of the correlations. We will also extend the Merton
model~\cite{merton74} of credit risk to account for the
non-stationarity. To this end, we will now put forward a
re-interpretation of the Wishart random matrix model for correlation
matrices~\cite{Schmitt2013}. As mentioned in Sec.~\ref{sec21}, the Wishart model in its
original and widely used form is based on the assumption of
stationarity. Using ergodicity, it predicts statistical properties of
large \textit{individual} correlation and covariance matrices with the 
help of a \textit{fictitious} ensemble of random matrices. We now
argue that the Wishart model may be viewed as an ensemble of random
matrices that models a \textit{truly existing ensemble} of
non-stationary covariance matrices. Two elements in this latter
ensemble are shown in Fig.~\ref{fig3}, the whole ensemble consists of
all correlation matrices measured with a window of length $T$ sliding
through a set of much longer time series of length $T_\textrm{tot}$.
The size of the truly existing ensemble is thus $T_\textrm{tot}/T$ if
the windows do not overlap. The average correlation or covariance
matrices $C_0$ or $\Sigma_0$ are simply the sample averages over the
whole time series of length $T_\textrm{tot}$. We have $K$ time series
divided in pieces of length $T$ that yield the truly existing
ensemble.  To model it with an ensemble of random matrices we have to
employ data matrices $A$ with $K$ rows, representing the model time
series, but we are free to choose their length $N$. As argued above,
the length of the time series controls the strength of the
fluctuations around the mean. Thus, we use $K\times N$ random data
matrices $A$ and write
\begin{equation}
w(A|\Sigma_0) \ = \ \frac{1}{\det^{N/2}(2\pi \Sigma_0)} 
             \exp\left(-\frac{1}{2}\text{tr} A^\dagger \Sigma_0^{-1} A \right) 
\label{wishart4}
\end{equation}
for the probability density function. The $K\times K$ mean covariance
matrix $\Sigma_0$ is the input and given by the sample mean using the
whole time series of length $T_\textrm{tot}$. This is our
re-interpreted Wishart model to describe fluctuating, non-stationary
covariance or correlation matrices. Importantly, ergodicity reasoning
is not evoked here, it would actually be wrong. It is also worth mentioning 
that we are not restricted to large matrix dimensions.

\begin{figure}[htbp]
  \begin{center}
    \includegraphics[width=0.55\textwidth]{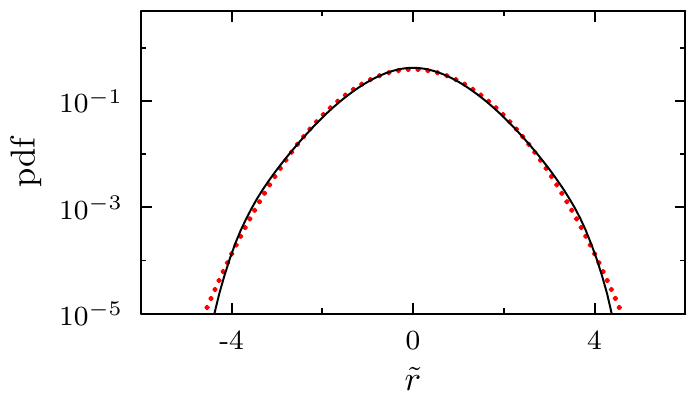}
  \end{center}
  \caption{Aggregated distribution of normalized returns $\tilde{r}$
    for fixed covariances from the S\&P 500 dataset, $\Delta t=1$ trading
    day and window length $T=25$ trading days. The circles show a
    normal distribution. Taken from~\cite{Schmitt2013}.}
 \label{fig4}
\end{figure}

Next, we demonstrate that the non-stationarity in the correlations
induces generic, \textit{i.e.},~universal features in financial time
series of correlated markets.  We begin with showing that the returns
are to a good approximation multivariate Gaussian distributed, if the
covariance matrix $\Sigma$ is fixed. We begin with assuming that the
distribution of the $K$ dimensional vectors
$r(t)=(r_1(t),\ldots,r_K(t))$ for a fixed return interval $\Delta t$
while $t$ is running through the dataset is given by
\begin{equation}
g(r|\Sigma) \ = \ \frac{1}{\sqrt{\det(2\pi\Sigma)}} 
                 \exp\left( -\frac{1}{2} r^\dagger \Sigma^{-1}r\right)  \ ,
\label{multivar}
\end{equation}
where we suppress the argument $t$ of $r$ in our notation. We test
this assumption with the daily S\&P 500 data. We divide the time series
in windows of length $T=25$ trading days which is short enough to
ensure that the sampled covariances can be viewed as constant within
these windows. We aggregate the data, \textit{i.e.},~we rotate
the return vector into the eigenbasis of $\Sigma$ and normalize
with the corresponding eigenvalues. As seen in Fig.~\ref{fig4}
there is good agreement with a Gaussian over at least four orders of
magnitude, details of the analysis can be found
in~\cite{Schmitt2013}. To account for the non-stationarity of the
covariance matrices, we replace them with random covariance matrices
\begin{equation}
\Sigma \quad \longrightarrow \quad \frac{1}{N}AA^\dagger \ ,
\label{random}
\end{equation}
drawn form the distribution \eqref{wishart4}. We emphasize that the
random matrices $A$ have dimension $K\times N$.  The larger $N$, the
more terms contribute to the individual matrix elements of
$AA^\dagger/N$, eventually fixing them for $N\to\infty$ to the mean
$\Sigma_0$. The fluctuating covariances alter the multivariate
Gaussian \eqref{multivar}. We model this by the ensemble
averaged return distribution
\begin{equation}
\langle g\rangle(r|\Sigma_0,N) \ = \ \int d[A]g\left(r\bigg|\frac{1}{N}AA^\dagger\right)
                                             w(A|\Sigma_0)  \ ,
\label{mvaver}
\end{equation}
which parametrically depends on the fixed empirical covariance matrix
$\Sigma_0$ as well as on $N$. The ensemble average can be done
analytically~\cite{Schmitt2013} and results in
\begin{eqnarray}
\langle  g \rangle (r|\Sigma_0,N) \ = \ 
                \frac{1}{2^{N/2+1}\Gamma(N/2)\sqrt{\det(2\pi\Sigma_0/N)}}
      \frac{\mathcal{K}_{(K-N)/2}\left(\sqrt{Nr^\dagger\Sigma_0^{-1}r}\right)}
                                      {\sqrt{Nr^\dagger\Sigma_0^{-1}r}^{(K-N)/2}} \ ,
\label{ergebnis}
\end{eqnarray}
where $\mathcal{K}_\nu$ is the modified Bessel function of the second
kind of order $\nu$.  In the data analysis below, we will find $K>N$.
Since the empirical covariance matrix $\Sigma_0$ is fixed, $N$ is the
only free parameter in the distribution \eqref{ergebnis}.  For large
$N$ it approaches a Gaussian. The smaller $N$, the heavier the tails,
for $N=2$ the distribution is exponential. Importantly, the returns
enter $\langle g \rangle (r|\Sigma_0,N)$ only via the bilinear form
$r^\dagger\Sigma^{-1}r$. 

\begin{figure}[htbp]
  \begin{center}
    \includegraphics[width=0.49\textwidth]{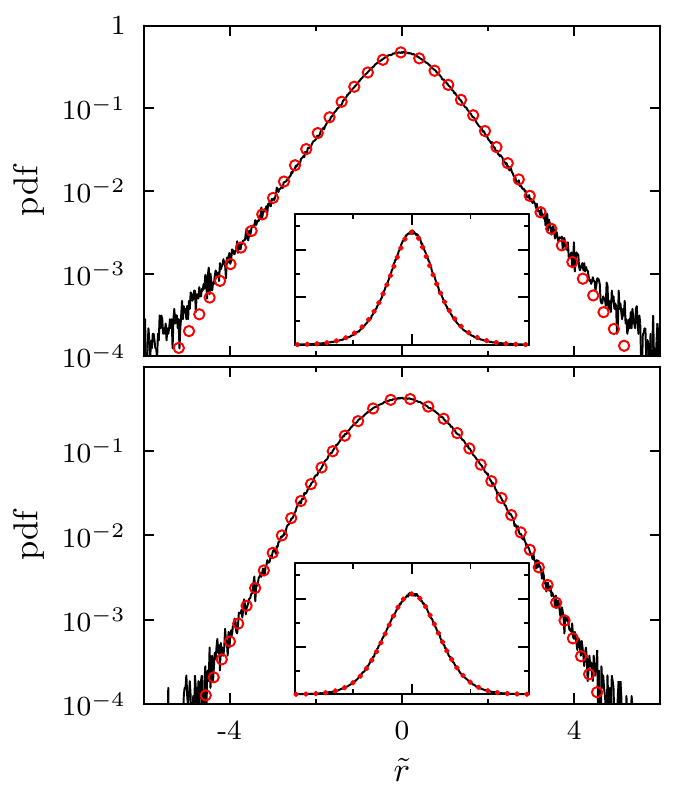}
  \end{center}
  \caption{Aggregated distribution of the rotated and scaled returns
    $\tilde{r}$ for $\Delta t =1$ (top) and $\Delta t =20$ (bottom)
    trading days. The circles correspond to the aggregation of the
    distribution~\eqref{ergebnis}. Taken from~\cite{Schmitt2013}.}
 \label{fig6}
\end{figure}

\begin{figure}[htp]
 \centering
 \includegraphics[width=0.75\textwidth]{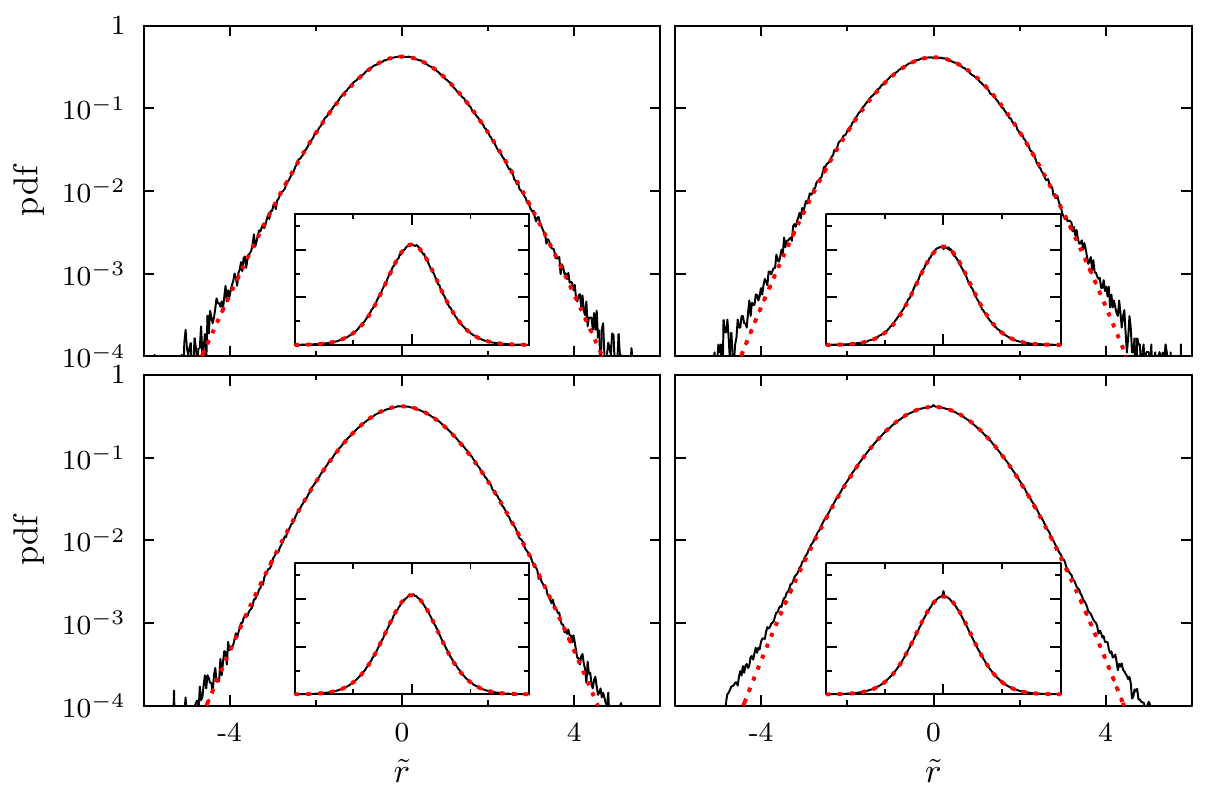}
  \caption{Aggregated distribution for the normalized monthly returns with empirical covariance matrix on a logarithmic scale. The black line shows the empirical distribution, the red dotted line shows the theoretical results. The insets show the corresponding linear plots. Top left/right: S\&P 500 (1992-2012) / (2002-2012), bottom left/right: NASDAQ (1992-2012) / (2002-2012). Taken from~\cite{Schmitt2015}.}
 \label{fig:Comp_EnsembleApproach_Heterog}
\end{figure}

To test our model, we again have to aggregate the data, but now for
the entire S\&P 500 dataset from 1992 to 2012, \textit{i.e.},~$T_\textrm{tot}=5275$
days, see Fig.~\ref{fig6}. We find $N=5$ for daily returns, \textit{i.e.}, $\Delta t =1$
trading day, and $N=14$ for $\Delta t =20$ trading days.
Furthermore, a more detailed comparison with larger datasets for monthly returns is provided in Fig.~\ref{fig:Comp_EnsembleApproach_Heterog}. Here, stocks taken from the S\&P 500 index and stocks taken from NASDAQ are used. In the top left corner the dataset consists of 307 stocks taken from S\&P 500 index which are continuously traded in the period 1992--2012. The other datasets following clockwise are: 439 stocks from S\&P 500 index in the time period 2002--2012, 2667 stocks from NASDAQ in the time period 2002--2012 and 708 stocks from NASDAQ in the time period 1992--2012. We find values around $N=20$ for monthly returns. Both datasets are available on \cite{yahoo}.
There is a good agreement between model and data.
Importantly, the distributions have heavy tails which result from the
fluctuations of the covariances, the smaller $N$, the heavier.  For
small $N$ there are deviations between theory and data in the tails.
Three remarks are in order. First, one should clearly distinguish this
multivariate analysis from the stylized facts of \textit{individual}
stocks which are well-known to have heavy-tailed distributions.
This is to some extent accounted for in our model, as seen in the
bottom part of Fig.~\ref{fig6}. In the top part the tails are heavier
because the time interval $\Delta t$ is much shorter. To further account
for this, we need to modify the Wishart model by using a distribution
different from a Gaussian~\cite{Meudt2015}.  Second, Fig.~\ref{fig3}
clearly shows that the empirical ensemble of correlation matrices has
inner structures which are also contained in our model, because the
mean $\Sigma_0$ enters.  Third, the consequences for portfolio
management are discussed in~\cite{chetalova}.

%%%%%%%%%%%%%%%%%%%%%%%%%%%%%%%%%%%%%%%%%%%%%%%%%%%%%%%%%%%%%%%%%%%%%%%%%%%%%%%%%%%%%%%%%%%

\section{Modeling Fluctuating Asset Correlations in Credit Risk}
\label{sec3}

Structural credit risk models employ the asset value at maturity to derive default events and their ensuing losses. Thus the distribution which describes the asset values have to be chosen carefully. One major requirement is that the distribution is in good accordance with empirical data. This goal can be achieved by using the random matrix approach for the asset correlations, discussed in Sec.~\ref{sec2}. Based on \cite{Schmitt2014,Schmitt2015}, we discuss the Merton model together with the random matrix approach in Sec.~\ref{sec31}. In Sec.~\ref{sec32} we reveal the results for the average loss distribution of a credit portfolio. The adjustability of the model is shown in Sec.~\ref{sec33}. In Sec.~\ref{sec34} we discuss the impact of the random matrix approach on VaR and ETL.

\subsection{Random Matrix Approach}
\label{sec31}
We start out from the Merton model \cite{merton74} and extend it considering a portfolio of $K$ credit contracts. Each obligor in the portfolio is assumed to be a publicly traded company. The basic idea is that the asset value $V_k(t)$ of company $k$ is the sum of time-independent liabilities $F_k$ and equity $E_k(t)$, \textit{i.e.},~$V_k(t)=F_k+E_k(t)$. $V_k(t)$ is considered as a stochastic process and in the spirit of the Merton model it is modeled by a geometric Brownian motion. Therefore one can trace back the changes in asset values to stock price returns and estimate the parameters of the stochastic process like volatility and drift by empirical stock price data. The liabilities mature after some time $T_\textrm{M}$ and the obligor has to fulfill his obligations and make a required payment. Thus, he has to pay back the face value $F_k$ without any coupon payments in between. This is related to a zero coupon bond and the equity of the company can be viewed as an European call option on its asset value with strike price $F_k$. A default occurs only if at maturity the asset value $V_k(T_\textrm{M})$ is below the face value $F_k$. The corresponding normalized loss is
\begin{equation}
L_k = \frac{F_k - V_k(T_\textrm{M})}{F_k} \:\Theta(F_k-V_k(T_\textrm{M}))\;.
\label{eq:losses}  
\end{equation}
The Heaviside step function $\Theta(x)$ guarantees that a loss is always larger than zero. This is necessary, because in the case $V_k(T_\textrm{M})>F_k$ the company is able to make the promised payment and no loss occurs. In other words, the default criterion can be affiliated to the leverage at maturity $F_k/V_k(T_\textrm{M})$. If the leverage is larger than one a default occurs and if the leverage is below one no default occurs. The total portfolio loss $L$ is a sum over the individual losses weighted by their fractions $f_k$ in the portfolio
\begin{equation}
L=\sum\limits^{K}_{k=1}{f_k L_k}\hspace{1cm},\hspace{1cm}f_k=\frac{F_k}{\sum^{K}_{i=1}F_i}\;.
\label{eq:Portfolio_Loss}
\end{equation}
The aim is to describe the average portfolio loss distribution $p(L)$ which can be expressed by means of a filter integral
\begin{equation}
p(L)=\int\limits_{[0,\infty)^K}d[V]g(V|\Sigma)\delta\left(L-\sum\limits^{K}_{k=1}{f_k L_k}\right)\;,
\label{eq:Loss_Convolution}
\end{equation}
where $g(V|\Sigma)$ is the multivariate distribution of all asset values at maturity time $T_\textrm{M}$ and $\Sigma$ is the covariance matrix. This is equivalent to a $K-1$ fold convolution which is expressed in terms of a filter integral by means of the Dirac delta function $\delta(x)$. We notice the complexity of the integral \eqref{eq:Loss_Convolution} as the losses \eqref{eq:losses} involve Heaviside functions. The distribution $g(V|\Sigma)$ is obtained by the more easily accessible distribution $g(r|\Sigma)$ where $r$ is the return vector consisting of the returns
\begin{equation}
r_k(t)=\frac{V_k(t+\Delta t)-V_k(t)}{V_k(t)}\;,
\end{equation}
defined analogously to \eqref{returns}. Here $\Delta t$ is the return horizon which corresponds with the maturity time, \textit{i.e.},
\begin{equation}
\Delta t=T_\textrm{M}
\end{equation}
because we are interested in changes of the asset values over the time period $T_\textrm{M}$.

The crucial problem is that the asset values show fluctuating correlations in the course of time. This non-stationarity has to be taken into account by the distribution $g(r|\Sigma)$ when larger time scales like one year or more are considered. As described in Sec.~\ref{sec2} the random matrix approach can be used to cope with the non-stationary asset correlations. The average asset value distribution $\left\langle g\right\rangle(V|\Sigma_0,N)$ is obtained by averaging a multivariate normal distribution over an ensemble of Wishart distributed correlation matrices. Thus, we calculate the loss distribution as an ensemble average. From \eqref{eq:Loss_Convolution} we find
\begin{equation}
\langle p\rangle(L|\Sigma_0,N)=\int\limits_{[0,\infty)^K}d[V]\langle g\rangle(V|\Sigma_0,N)\delta\left(L-\sum\limits^{K}_{k=1}{f_k L_k}\right)\;.
\label{eq:AverageLossDist}
\end{equation}
Again, we emphasize that the ensemble truly exists as a consequence of the non-stationarity. As a side effect of the random matrix approach the resulting distribution depends only on two parameters. The $K\times K$ average covariance matrix $\Sigma$ and the free parameter $N$ which controls the strength of the fluctuations around the average covariance matrix. $N$ behaves like an inverse variance of the fluctuations, the smaller $N$ the larger the fluctuations become. Both parameters have to be determined by historical stock price data.

\begin{figure}[htbp]
  \begin{center}
    \includegraphics[width=0.75\textwidth]{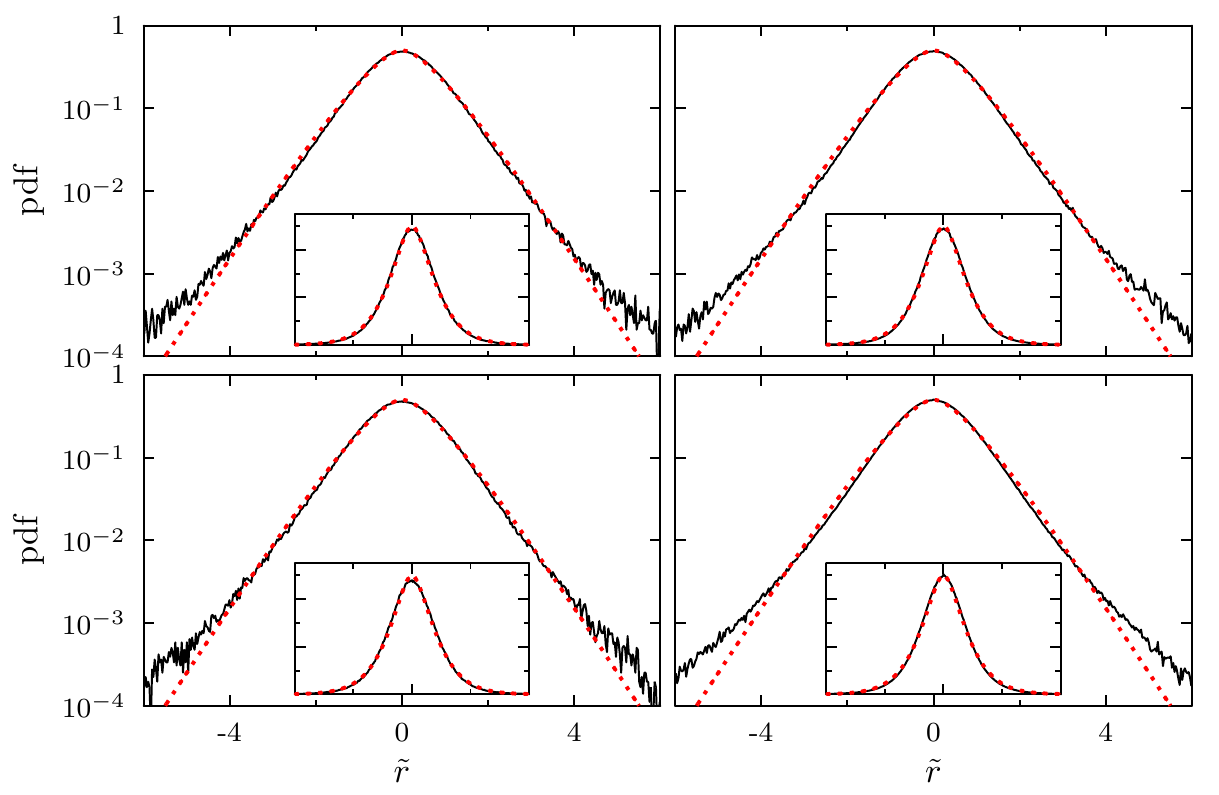}
  \end{center}
 \caption{Aggregated distribution for the normalized monthly returns with effective correlation matrix on a logarithmic scale. The black line shows the empirical distribution, the red dotted line shows the theoretical results. The insets show the corresponding linear plots. Top left/right: S\&P 500 (1992-2012) / (2002-2012), bottom left/right: NASDAQ (1992-2012) / (2002-2012). The average correlation coefficients are $c=0.26$, 0.35, 0.21 and $0.25$, respectively. Taken from~\cite{Schmitt2015}.}
 \label{fig:Comp_EnsembleApproach_Homog}
\end{figure}

The average asset value distribution depends on the $K\times K$ mean covariance matrix $\Sigma_0$. To circumvent the ensuing complexity and to make analytical progress we assume an effective average correlation matrix
\begin{align}
C_0=\begin{bmatrix}
	1 & c & c & \dots\\
	c & 1 & c & \dots\\
	c & c & 1 & \dots\\
	\vdots & \vdots & \vdots & \ddots
\end{bmatrix}
\label{eq avcorrm}
\end{align}
where all off-diagonal elements are equal to $c$. The average correlation is calculated over all assets for the selected time horizon. We emphasize that only the effective average correlation matrix $C_0$ is fixed, the correlations in the random matrix approach fluctuate around this mean value. In the sequel, whenever we mention a covariance matrix with effective correlation matrix we denote it as effective covariance matrix and whenever we mention a fully empirical covariance matrix where all off-diagonal elements differ from another we denote it as empirical covariance matrix or covariance matrix with heterogeneous correlation structure. Using the assumption \eqref{eq avcorrm}, analytical tractability is achieved but it also raises the question whether the data can still be described well. To compare the result with data one has to rotate and scale the returns again, but instead of using the empirical covariance matrix the covariance matrix with effective average correlation structure has to be applied. The results for monthly returns, using the same dataset as in Fig.~\ref{fig:Comp_EnsembleApproach_Heterog}, are shown in Fig.~\ref{fig:Comp_EnsembleApproach_Homog}.
Still, there is a good agreement between the average asset value distribution with assumption \eqref{eq avcorrm} and the data. This leads to the conclusion that the approximation is reasonable. Considering the parameter $N_\textrm{eff}$, which is needed to describe the fluctuations around the effective average correlation matrix, values around $N_\textrm{eff}=4$ are found. In contrast to the larger values around $N=20$ which describe the distributions best in case of an empirical correlation matrix the lower values in case of an effective correlation matrix with average correlation $c$ are needed to describe the larger fluctuations around this average. This result corroborates the interpretation of $N$ as an inverse variance of the fluctuations. Now the correlation structure of a financial market is captured by solely two parameters. The average correlation coefficient $c$ and parameter $N$ which indicates the strength of the fluctuations around this average.

\subsection{Average Loss Distribution}
\label{sec32}
Having shown the quality of the random matrix approach we may now proceed in calculating the average portfolio loss distribution \eqref{eq:AverageLossDist}. We deduce the average distribution for the asset values $\langle g\rangle(V|\Sigma_0,N)$ from the result \eqref{ergebnis} for the returns. In the Merton model it is assumed that the asset values $V_k(t)$ follow a geometric Brownian motion with drift and volatility constants $\mu_k$ and $\rho_k$, respectively. This leads to a multivariate Gaussian of the form \eqref{multivar} for the returns, which is consistent with the random matrix approach. Therefore, according to It\^o's Lemma \cite{Ito}, we perform a change of variables
\begin{align}
 r_k \longrightarrow \ln\frac{V_k(T_\textrm{M})}{V_{k0}} - \left(\mu_k-\frac{\rho_k^2}{2}\right)T_\textrm{M}\;,
\end{align}
with $V_{k0}=V_k(0)$ and the volatilities
\begin{align}
 \rho_k=\frac{\sigma_k}{\sqrt{T_\textrm{M}}}\;,
\end{align}
where $\sigma_k$ is the standard deviation in connection with \eqref{corr}. Furthermore we assume a large portfolio in which all face values $F_k$ are of the same order and carry out an expansion for large $K$. The analytical result is
\begin{align}
\langle p \rangle (L|c,N) & = \frac{ 1 }{ \sqrt{ 2 \pi } 2^{N/2} \Gamma( N / 2 ) } \int_0^\infty dz \ z^{N/2 -1} \text{e}^{-z/2}   \sqrt{ \frac{ N }{ 2 \pi } }\notag \\
& \qquad \times \int_{-\infty}^{+\infty} du \exp\left(-\frac{ N }{ 2 } u^2 \right) \frac{ 1 }{ \sqrt{ M_2(z,u) } } \exp\left( - \frac{ ( L - M_1(z,u) )^2 }{ 2 M_2(z,u) } \right)
\label{eq:LossDist}
\end{align}
for the average loss distribution with
\begin{align}
M_1(z,u) =  \sum_{k=1}^K f_k m_{1k}(z,u)
\label{eq:M1}
\end{align}
and
\begin{align}
 \quad M_2(z,u) = \sum_{k=1}^K f_k^2 \left( m_{2k}(z,u) - m_{1k}^2(z,u) \right) \;.
 \label{eq:M2}
\end{align}
The $j$-th moments $m_{jk}(z,u)$ are

\begin{align}m_{jk}(z,u) & = \frac{ \sqrt{N} }{ \rho_k \sqrt{ 2 \pi T_\textrm{M} ( 1 - c ) } }  \int_{-\infty}^{\hat{F}_k} d \hat{V}_k \ \left( 1 - \frac{ V_{k0} }{ F_k } \exp\left( \sqrt{z} \hat{V}_k +  \left( \mu_k - \frac{ \rho_k^2 }{ 2 } \right)T_\textrm{M} \right) \right)^j \notag \\ 
& \qquad \times \exp\left(- \frac{ \left( \hat{V}_k + \sqrt{c T_\textrm{M}} u \rho_k \right)^2 }{ 2 T_\textrm{M} ( 1 - c) \rho_k^2 / N }  \right)\;,
\label{eq:mjk}
\end{align} 
see \cite{Schmitt2014}. The changed variable is $\hat{V}_k=(\ln(V_k(T_\textrm{M})/V_{k0}) - (\mu_k-\rho_k^2/2)T_\textrm{M})/\sqrt{z}$ with the upper bound for the integral \eqref{eq:mjk}
\begin{align}
	\hat{F}_k & = \frac{ 1 }{ \sqrt{z} } \left(\ln \frac{ F_k }{ V_{k0} } - \left( \mu_k - \frac{ \rho_k^2 }{ 2 } \right) T_\textrm{M} \right)  \;.
\end{align}
The integrals in \eqref{eq:LossDist} have to be evaluated numerically.

To further illustrate the results, we assume homogeneous credit portfolios. A portfolio is said to be homogeneous when all contracts have the same face value $F_k=F$ and start value $V_k(0)=V_0$ and the same parameters for the underlying stochastic processes like volatility $\rho_k=\rho$ and drift $\mu_k=\mu$. Of course this does not mean that all asset values follow the same path from $t=0$ to maturity $T_\textrm{M}$ because underlying processes are stochastic.

\begin{figure}[htbp]
  \begin{center}
    \includegraphics[width=0.85\textwidth]{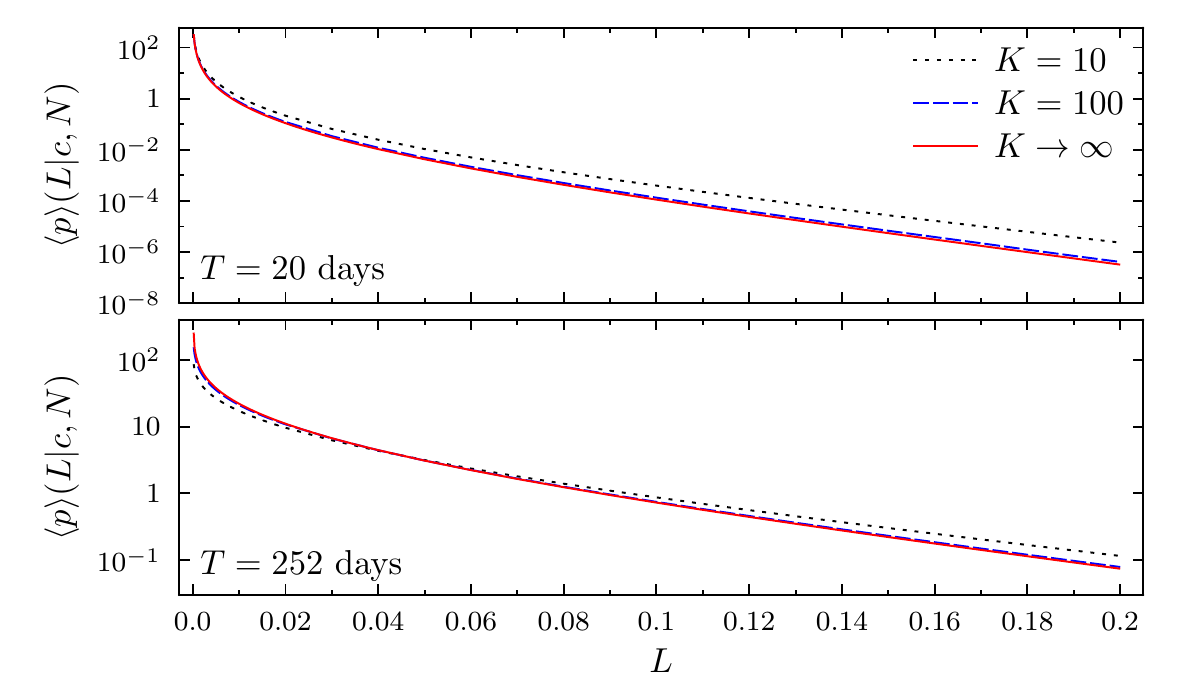}
  \end{center}
 \caption{Average portfolio loss distribution for different portfolio sizes of $K=10$, $K=100$ and the limiting case $K\to\infty$. At the top the maturity time is one month, at the bottom it is one year. Taken from~\cite{Schmitt2015}.}
 \label{fig:Loss_Distr_Corr}
\end{figure}

It is often argued that diversification significantly reduces the risk in a credit portfolio. In the context mentioned here, diversification solely means the increase of the number $K$ of credit contracts in the credit portfolio on the same market. The limit distribution for an infinitely large portfolio provides information whether this argument is right or wrong. We thus consider a portfolio of size $K\to\infty$ and find the limiting distribution
\begin{align}
\langle p \rangle(L|c,N)\bigg|_{K\to\infty} & =  \frac{ 1 }{ 2^{N/2} \Gamma( N/2) } \sqrt{ \frac{ N }{ 2 \pi } } \int_0^\infty dz \, z^{N/2-1} e^{-z/2} \exp\left(-\frac{ N }{ 2 } u_0^2\right) \frac{ 1 }{  \left| \partial m_1 (z,u) / \partial u \right|_{z,u_0} }\;,
\end{align}
where $u_0(L,z)$ is the implicit solution of the equation
\begin{align}
L & = m_1(z,u_0)\;.
\label{eq:fzu}
\end{align}
We now display the average loss distribution for different $K$. The model depends on four parameters which can be calibrated by empirical data. Three of them, the average drift $\mu$, the average volatility $\rho$ and the average correlation coefficient $c$ can be directly calculated from the data. The fourth parameter $N$, controlling the strength of the fluctuations, has to be determined by fitting the average asset value distribution onto the data. The resulting average portfolio loss distribution $\langle p\rangle(L|c,N)$ for correlation averaged asset values is shown in Fig.~\ref{fig:Loss_Distr_Corr}.
Different portfolio sizes $K=10,100$ and $K\to\infty$ and two different maturity times $T_\textrm{M}=20$ trading days and $T_\textrm{M}=252$ trading days are shown. For the estimation of the empirical parameters the S\&P 500 dataset in the time period 1992--2012 is used. The parameters for $T_\textrm{M}=20$ trading days are $N=4.2$, $\mu = 0.013$ month$^{-1}$, $\rho=0.1$ month$^{-1/2}$ and an average correlation coefficient of $c=0.26$, shown on the top, and for a maturity time of $T_\textrm{M}=1$ year $N=6.0$, $\mu = 0.17$ year$^{-1}$, $\rho=0.35$ year$^{-1/2}$ and an average correlation coefficient of $c=0.28$, shown on the bottom. Moreover a face value of $F=75$ and an initial asset value of $V_0=100$ is used. There is always a slowly decreasing heavy-tail. A significant decrease of the risk of large losses cannot be achieved by increasing the size of the credit portfolio. Instead the distribution quickly converges to the limiting distribution $K\to\infty$. This drastically reduces the effect of diversification. In a quantitative manner it is thus shown that diversification does not work for credit portfolios with correlated asset values. Speaking pictorially, the correlations glue the obligors together and let them act to some extent like just one obligor.

The values of the average correlation coefficient $c$ and the parameter $N$ also influence the average loss distribution. The larger the average correlation $c$ and the smaller the parameter $N$, the heavier are the tails of the distribution and the more likely is the risk of large losses. 

\subsection{Adjusting to Different Market Situations}
\label{sec33}
\begin{figure}[htbp]
  \begin{center}
    \includegraphics[width=0.85\textwidth]{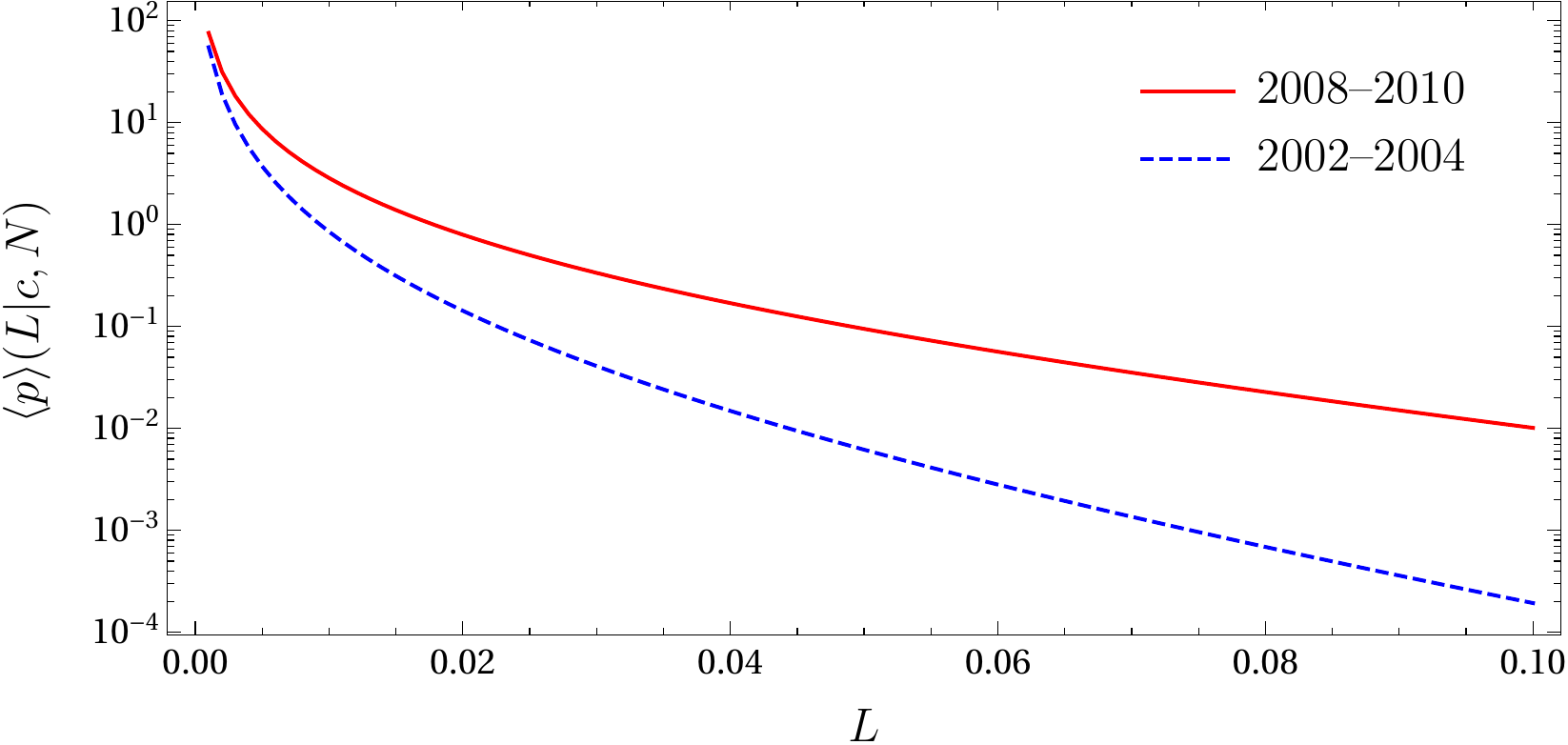}
  \end{center}
 \caption{Average loss distribution for different parameters taken from Tab.~\ref{tab1}. The dashed line corresponds to the calm period 2002--2004, the solid line corresponds to the global financial crisis 2008--2010.}
 \label{fig:Comp}
\end{figure}

The non-stationarity of financial markets implies that there are calm periods where the markets are stable, as well as periods of crisis as in the period 2008--2010, see, \textit{e.g.},~for the volatility in Fig.~\ref{fig1}. Observables describing the market behavior in different periods vary significantly. Consequently, the loss distribution, particularly its tail, strongly changes in different market situations. Our model fully grasps this effect. The parameters, \textit{i.e.},~drift, volatility, average correlation coefficient and parameter $N$ can be adjusted to different periods. To demonstrate the adjustability of our model based on the random matrix approach we consider the two periods 2002--2004 and 2008--2010. The first period is rather calm whereas the second includes the global financial crisis. We determine the average parameters for monthly returns of continuously traded S\&P 500 stocks, shown in Tab.~\ref{tab1}.
\begin{table}[htbp]
\centering
\begin{tabular}{ccccccc}%{rrrrrrr}
\hline
Time horizon & $K$ & $N_\textrm{eff}$ &  $ \rho$ in & $ \mu$ in & $c$ \\ 
for estimation &   &  & month$^{-1/2}$ & month$^{-1}$  &  & \\ 
\hline
2002-2004 & 436 & 5 & 0.10 & 0.015 & 0.30\\
2008-2010 & 478 & 5 & 0.12 & 0.01 & 0.46 \\
\hline
\end{tabular}
\caption{Average parameters used for two different time horizons. Taken from \cite{Schmitt2015}.}
\label{tab1}
\end{table}
For each period we take the corresponding parameters and calculate the average portfolio loss distribution, see Fig.~\ref{fig:Comp}.
As expected we find a much more pronounced tail risk in times of crisis. This is mainly due to the enlarged average correlation coefficient in times of crisis. Consequently we are able to adjust the model onto various periods. It even is possible to adjust the parameters and hence the tail behavior dynamically.

The setting discussed here includes avalanche or contagion effects only indirectly when calibrated to a market situation in the state of crisis. Direct modeling of contagion is provided in \cite{hatchett,kuehn}.

\subsection{Value at Risk and Expected Tail Loss}
\label{sec34}
The approximation of an effective correlation matrix facilitated analytical progress, but importantly the average asset return distribution still fits empirical data well when using this approximation. We now show that this approximation is also capable of estimating the value at risk (VaR) and the expected tail loss (ETL), also referred to as expected shortfall. We compare the results obtained in this approximation with the results obtained for an empirical covariance matrix. This is also interesting from risk management because it is common to estimate the covariance matrix over a long period of time and use it as an input for various risk estimation method. Put differently, we are interested in the quality of risk estimation using an effective correlation matrix and taking fluctuating correlations into account. 

The comparison of the effective correlation matrix with the empirical covariance matrix cannot be done analytically. Hence, Monte-Carlo simulations to calculate the VaR and ETL are carried out. For each asset its value at maturity time $T_\textrm{M}$ is simulated and the portfolio loss according to \eqref{eq:Portfolio_Loss} is calculated. All assets have the same fraction in the portfolio. For different time horizons the empirical covariance matrix, volatilities and drifts for monthly returns of the S\&P 500 stocks are calculated. In addition the parameter $N$ is determined as described above. In the calm period 2002--2004 we find for the empirical covariance matrix a rather large parameter value of $N=14$ whereas during the financial crisis 2008--2010 we find $N=7$. This once more illustrates the meaning of $N$ as an inverted variance of the fluctuations. 

The relative deviations of the VaR and ETL for different quantiles of the effective covariance matrix from the empirical covariance matrix are calculated. This is done in two different ways. First, one may assume a fully homogeneous portfolio where the average values for volatility and drift for each stock is used. Second, one may use the empirically obtained values for each stock. It turns out that in most cases the effective covariance matrix together with homogeneous volatility and drift underestimates the risk. In contrast, if one uses heterogeneous volatilities and drifts and the effective covariance matrix one finds a satisfactory agreement compared to the full empirical covariance matrix, see \cite{Schmitt2015}. In the latter case the effective covariance matrix slightly overestimates the VaR and ETL in most cases. Hence, the structure of the correlation matrix does not play a decisive role for the risk estimation. This is so because the loss distribution is always a multiply averaged quantity. A good estimation of the volatilities, however, is crucial.

\begin{figure}[htbp]
  \begin{center}
    \includegraphics[width=0.85\textwidth]{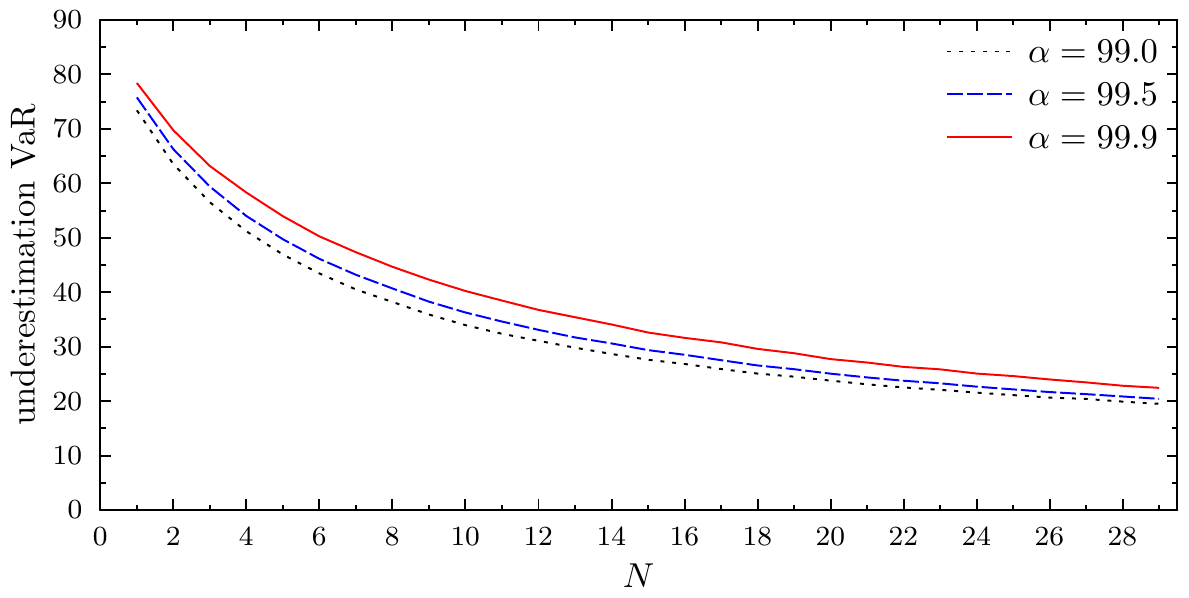}
  \end{center}
 \caption{Underestimation of the VaR if fluctuating asset correlations are not taken into account. The empirical covariance matrix is used and compared for different values of $N$. Taken from~\cite{Schmitt2015}.}
 \label{fig:VaR}
\end{figure}

The benefit of the random matrix approach is shown by comparing the VaR calculated for $N\to\infty$ and for different values of $N$. The case $N\to\infty$ does not allow fluctuations of the covariance matrix. This means that we use stationary correlations which turn the distribution $\langle g\rangle(V|\Sigma_0,N)$ of the asset values at maturity into a multivariate log-normal distribution. Thus, for $N\to\infty$ the benefits of the random matrix approach are disabled. The underestimation of the VaR by using stationary correlations, \textit{i.e.},~$N\to\infty$, is measured in terms of the relative deviation from the VaR calculated for empirical values of $N$. The empirical covariance matrix and the empirical, \textit{i.e.},~heterogeneous, volatilities and drifts calculated in the period 2006--2010 are used. The results are shown in Fig.~\ref{fig:VaR}.
Here, different quantiles $\alpha=0.99,0.995,0.999$ are used. For the empirically observed parameter $N=12$ the VaR is underestimated between 30\% and 40\%. Hence, to avoid a massive underestimation of risk, the fluctuations of the asset correlations must be accounted for.

\section{Concurrent Credit Portfolio Losses}
\label{sec4}

In the previous section solely one single portfolio on a financial market was considered. Here, based on \cite{Sicking2018}, we consider the problem of concurrent portfolio losses where two non-overlapping credit portfolios are taken into account. In Sec.~\ref{sec41} we discuss copulas of homogeneous portfolios. The dependence of empirical S\&P 500- and Nikkei-based credit portfolios is discussed in Sec.~\ref{sec42}.

\subsection{Simulation Setup}
\label{sec41}
\begin{figure}[htp]
 \centering
 \includegraphics[width=0.5\textwidth]{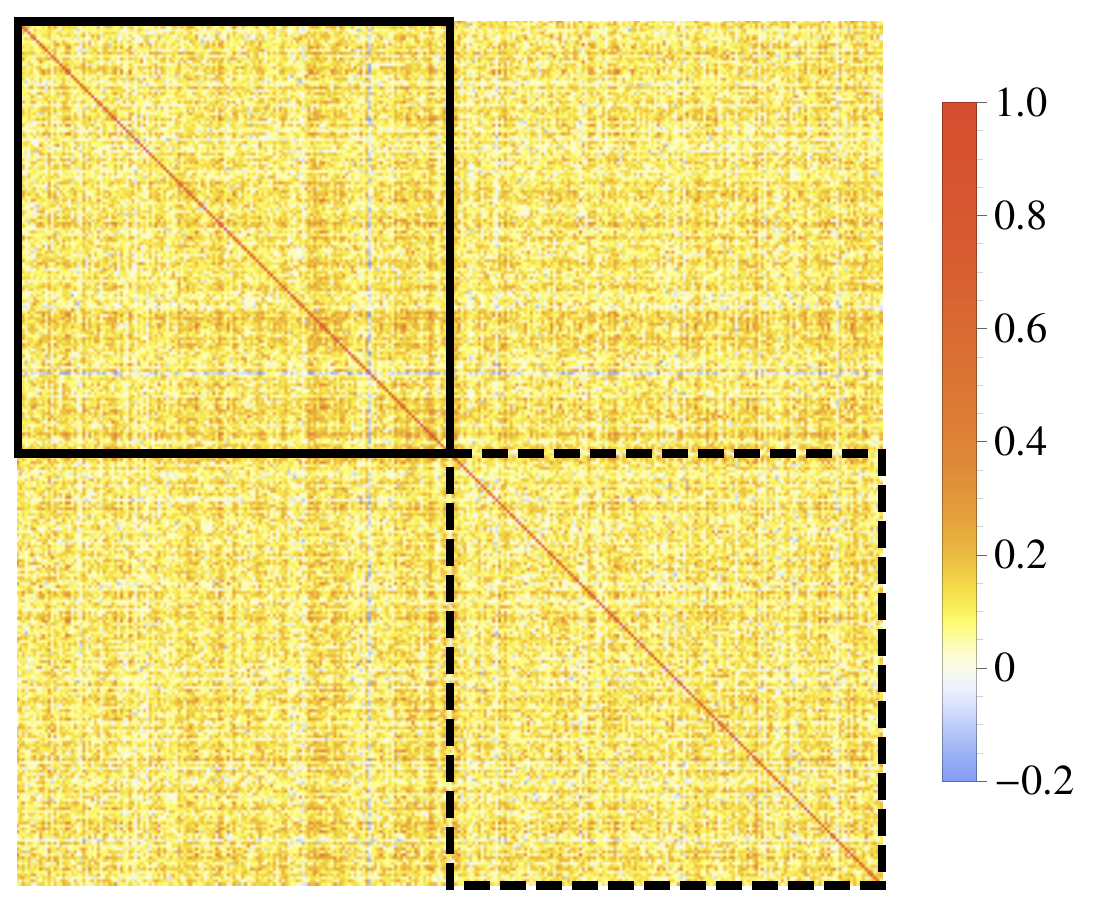}
  \caption{Heterogeneous correlation matrix illustrating a financial market. The two rimmed squares correspond to two non-overlapping credit portfolios. Taken from~\cite{Sicking2018}.}
 \label{fig:Two_Portfolios_Schematic}
\end{figure}

We consider two non-overlapping credit portfolios which are set up according to Fig.~\ref{fig:Two_Portfolios_Schematic}, in which the financial market is illustrated by means of its correlation matrix. 
The color indicates the strength of the correlation of two companies in the market. Hence, the diagonal is red as the diagonal of a correlation matrix is one by definition. The two portfolios are marked in Fig.~\ref{fig:Two_Portfolios_Schematic} as black rimmed squares. Both portfolios include $K$ contracts which means they are of equal size and no credit contract is contained in both portfolios. Despite the fact that the portfolios are non-overlapping they are correlated due to non-zero correlations in the off-diagonal squares.

The joint bivariate distribution of the losses $L^{(1)}$ and $L^{(2)}$ of two credit portfolios 
\begin{equation}
p(L^{(1)},L^{(2)})=\int\limits_{[0,\infty)^K}d[V]g(V|\Sigma)\delta\left(L^{(1)}-\sum\limits^{K}_{k=1}{f^{(1)}_k L^{(1)}_k}\right)\delta\left(L^{(2)}-\sum\limits^{K}_{k=1}{f^{(2)}_k L^{(2)}_k}\right)
\end{equation}
is defined analogously to \eqref{eq:Loss_Convolution}. Here, the upper index indicates the corresponding portfolio and the normalized losses $L^{(b)}_k$ and portfolio losses $L^{(b)}$ as well as the fractions $f^{(b)}_k$ for $b=1,2$ are defined analogously to \eqref{eq:losses} and \eqref{eq:Portfolio_Loss}, respectively. The total face value $F_k=F^{(1)}_k+F^{(2)}_k$ is the sum over the face values for both portfolios. We remark that for two non-overlapping portfolios one of the addends is always zero.

With this simulation setup the correlated asset values $V_k(T_\textrm{M})$ for each contract are simulated several thousand times to calculate the portfolio losses and out of them the empirical portfolio loss copula. A copula is the joint distribution of a bivariate random variable expressed as function of the quantiles for the two marginal distributions. The basic idea of copulas is to separate the mutual dependence of a bivariate random variable from the two marginal distributions to analyze the statistical dependencies. In particular we will analyze the copula density which is illustrated by means of a normalized two-dimensional histogram. Hence, when speaking of a copula, we rather mean its density. To obtain a better understanding of the mutual dependencies which are expressed by the empirical copula it is compared to a Gaussian copula. This Gaussian copula is fully determined by the correlation coefficient of the portfolio losses. 

\begin{figure}[htp]
 \centering
 \includegraphics[width=0.5\textwidth]{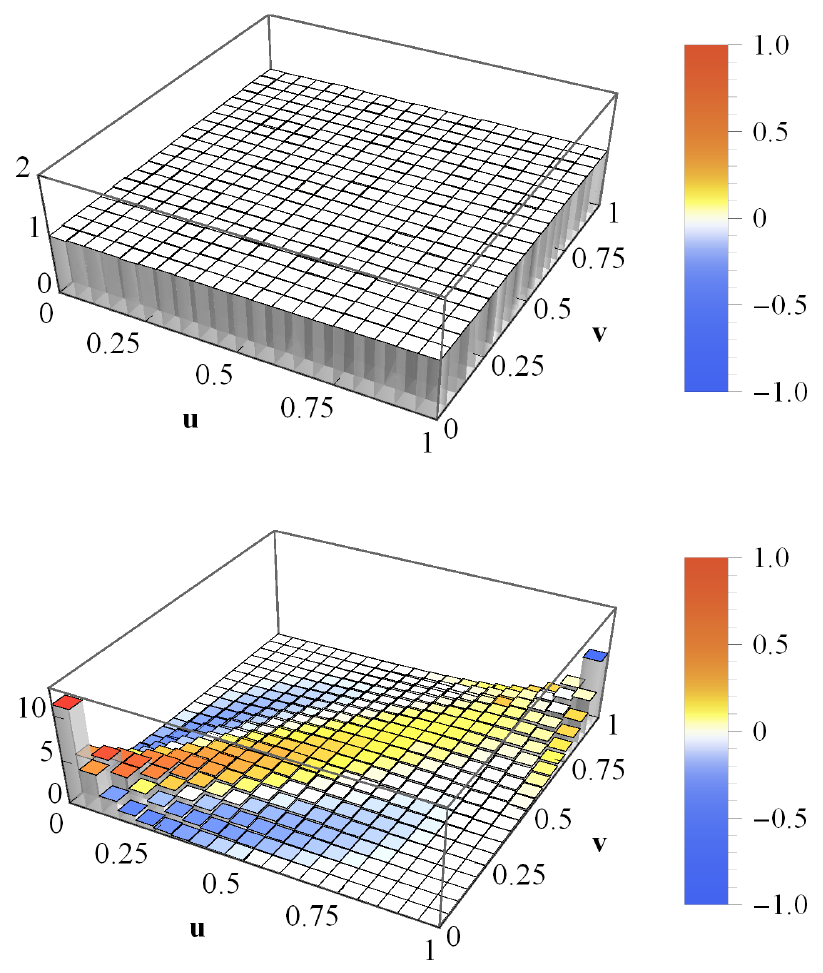}
  \caption{Average loss copula histograms for homogeneous portfolios with vanishing average asset correlations $c=0$. The asset values are multivariate log-normal ($N\to\infty$) in the top figure and multivariate heavy-tailed ($N_\textrm{eff}=5$) in the bottom figure. The color bar indicates the local deviations from the corresponding Gaussian copula. Taken from~\cite{Sicking2018}.}
 \label{fig:Copulas_Homogeneous_c=0_Differentns_cropped}
\end{figure}

To systematically study the influence of different parameters on the portfolio loss copula it is helpful to analyze homogeneous portfolios first. The most generic features can be found by focusing on asset correlations and drifts. The simulation is run in two different ways. First, we consider Gaussian dynamics for the stock price returns. This means that the asset values at maturity time $T_\textrm{M}$ are distributed according to a multivariate log-normal distribution. We notice that in the case of Gaussian dynamics the fluctuations of the random correlations around the average correlation coefficient is zero. This corresponds to the case $N\to\infty$. Second, we use fluctuating asset correlations, employing a parameter value of $N_\textrm{eff}=5$ in accordance with the findings of \cite{Schmitt2015} for an effective correlation matrix, see Tab.~\ref{tab1}. For the simulation the parameters $\mu = 10^{-3}\:\text{day}^{-1}$, $\rho = 0.03\:\text{day}^{-1/2}$ and leverages $F/V_0 = 0.75$ are chosen. The portfolios are of size $K=50$, the maturity time is $T_\textrm{M}=1\:\text{year}$ and a market with vanishing asset correlation, \textit{i.e.},~$c=0$, is considered. The resulting copulas are shown in Fig.~\ref{fig:Copulas_Homogeneous_c=0_Differentns_cropped}.
For $N\to\infty$ the loss copula is constant. This result is quite obvious. Due to the Gaussian dynamics and $c=0$ the asset values are uncorrelated and statistically independent. Therefore the portfolio losses, which are derived from those independent quantities, do not show mutual dependencies either. The resulting copula is an independence copula which agrees with a Gaussian loss copula for a portfolio loss correlation of $\mathrm{Corr}(L_1,L_2)=0$. In the color coding only white appears. The difference of the empirical copula and the Gaussian copula within each bin is illustrated by means of a coloring code. The color bar on the right hand side indicates the difference between the two copulas. The colors yellow to red imply a stronger dependence by the empirical copula in the given $(u,v)$-interval than predicted by the Gaussian copula. The colors turquoise to blue imply a weaker dependence of the empirical copula than by a Gaussian copula. Color white implies that the local dependence is equal. The empirical average loss correlation calculated from the simulation outcomes is zero and corroborates this result.

In the bottom panel of Fig.~\ref{fig:Copulas_Homogeneous_c=0_Differentns_cropped} the combination of $c=0$ and $N_\textrm{eff}=5$ is shown. The deviations from the independence copula are striking. They emerge because we included according to the random matrix approach fluctuating asset correlations around the average correlation $c=0$. In that way positive as well as negative correlations are equally likely. Having a look at the copula histograms we find a significant deviation from a Gaussian copula. A Gaussian copula is always symmetric regarding the line spanned from $(0,1)$ to $(1,0)$. Nevertheless, the portfolio loss correlation is $\mathrm{Corr}(L_1,L_2)=0.752$. The deviations from the Gaussian copula which is determined by the calculated correlation coefficient can be seen in Fig.~\ref{fig:Copulas_Homogeneous_c=0_Differentns_cropped}. Especially in the $(1,1)$ corner which is related to concurrent extreme losses we see that the empirical copula shows a weaker dependence than the Gaussian copula. We still have to answer the question why the portfolio losses exhibit such a strong positive correlation although the average asset correlation is set to zero in the simulation. First, as explained above, credit risk is highly asymmetric. For example, if in a credit portfolio one single contract generates a loss it is already sufficient enough that the whole portfolio generates a loss. The company defaulting may just cause a small portfolio loss but still it dominates all other non-defaulting and maybe prospering companies. In other words there is no positive impact of non-defaulting companies on the portfolio losses. All those non-defaults are projected onto zero. Second, the fluctuating asset correlations imply a division of the companies into two blocks. The companies show positive correlations within the blocks and negative correlations across them. Due to the aforementioned fact that non-defaulting companies have no positive impact on the loss distribution, the anti-correlations contribute to the portfolio loss correlation in a limited fashion. They would act as a risk reduction which is limited according to the asymmetry of credit risk. On the other side positive correlations within the blocks imply a high risk of concurrent defaults.

We now investigate the impact of the drift. All non-defaulting companies are projected onto a portfolio loss equal zero. The influence of these projections onto zero and therefore the default-non-default ratio can be analyzed in greater detail by varying the drift of the asset values. For example, if a strong negative drift is chosen it is highly likely that all companies will default at maturity. 

\begin{figure}[htp]
 \centering
 \includegraphics[width=0.5\textwidth]{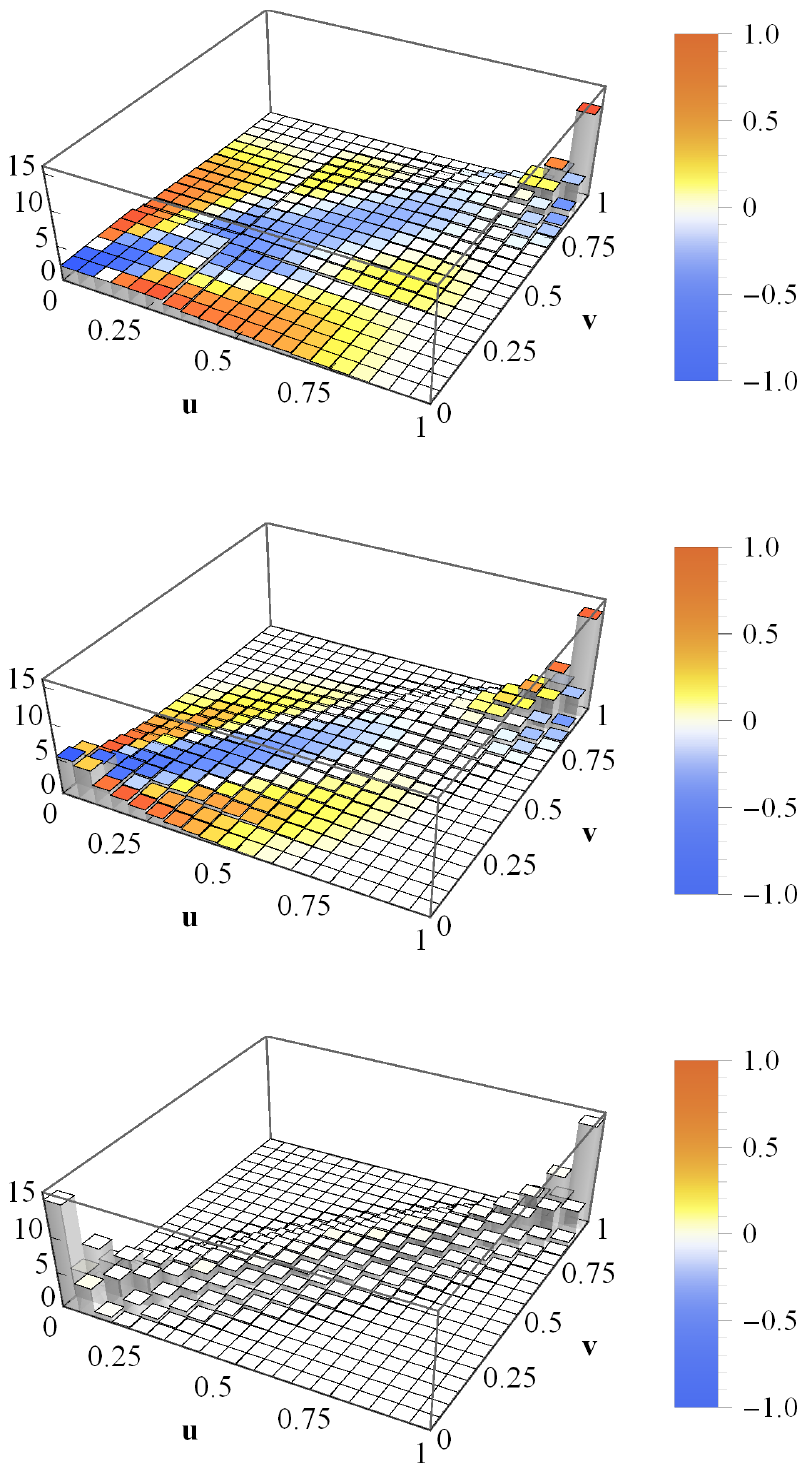}
  \caption{Average loss copula histograms for homogeneous portfolios with asset correlations $c=0.3$. The asset values are multivariate log-normal ($N\to\infty$). The drifts are ${\mu = 10^{-3}\:\text{day}^{-1}}$ (top), $3\times10^{-4}\:\text{day}^{-1}$ (middle) and ${-3\times10^{-3}\:\text{day}^{-1}}$ (bottom). The color bar indicates the local deviations from the corresponding Gaussian copula. Taken from~\cite{Sicking2018}.}
 \label{fig:Copulas_Homogeneous_DifferentMus_cropped}
\end{figure}

We consider Gaussian dynamics with an average asset correlation of $c=0.3$ and a volatility of $\rho = 0.02\:\text{day}^{-1/2}$ and different values of $\mu$. Fig.~\ref{fig:Copulas_Homogeneous_DifferentMus_cropped} shows the resulting copulas for three different drift parameters.
On the top panel a drift of $\mu = 10^{-3}\:\text{day}^{-1}$ was chosen, which leads to a non-default ratio of $39.1\%$ and an estimated portfolio loss correlation of $\mathrm{Corr}(L_1,L_2)=0.851$. One finds a significant deviation from a symmetric Gaussian copula. On the middle and bottom panel a drift of $\mu = 3\times 10^{-4}\:\text{day}^{-1}$ and $\mu = -3\times 10^{-3}\:\text{day}^{-1}$ was chosen, which leads to non-default ratios of $12.8\%$ and zero, respectively. The estimated portfolio loss correlations increase as the non-default ratios decrease, one find a correlation of $\mathrm{Corr}(L_1,L_2)=0.904$ and $\mathrm{Corr}(L_1,L_2)=0.954$, respectively. Moreover we see that the empirical copula turns ever more Gaussian if the percentage of non-defaults decreases. Finally at a default probability of $100\%$ the empirical loss copula is a Gaussian copula. This is seen in the bottom panel where no color except for white appears. In the middle and top panel we see deviations from the Gaussian copula. Especially in the $(1,1)$ corner we see that the empirical copula exhibits a stronger dependence than predicted by the corresponding Gaussian copula. In both cases the statistical dependence of large concurrent portfolio losses are underestimated by the Gaussian copula. 

We infer that an increase in default probability yields an increase in portfolio loss correlation. In addition we conclude that the loss of information, which is caused by the projections onto zero, is responsible for the observed deviations of the statistical dependencies from Gaussian copulas.

\subsection{Empirical Credit Portfolios}
\label{sec42}
Now more realistic portfolios with heterogeneous parameters are considered. To systematically study the influence of heterogeneity only the volatility is initially chosen to be heterogeneous. Afterwards we will proceed with the analysis of fully heterogeneous portfolios. The empirical parameters like asset correlation, drift and volatility are determined by historical datasets from S\&P~500 and Nikkei~225. 

In order to avoid any effect due to a specific parameter choice, the average over thousands of simulations run with different parameter values is calculated.

\begin{figure}[htp]
 \centering
 \includegraphics[width=0.5\textwidth]{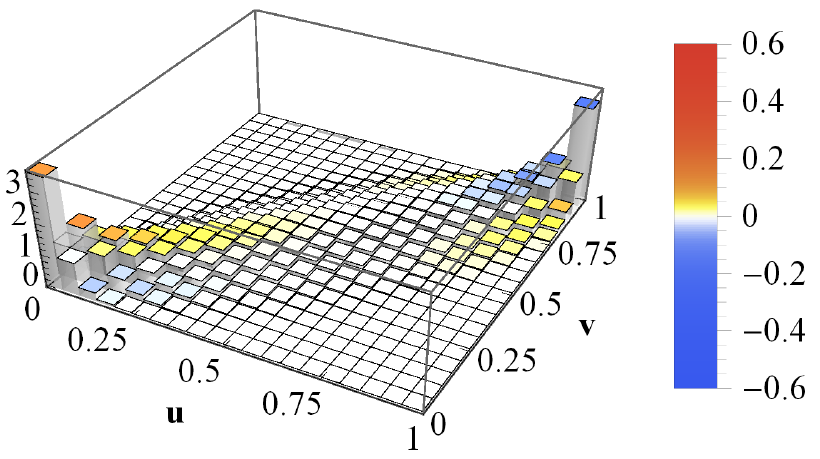}
  \caption{Average loss copula histograms for two portfolios with heterogeneous volatilities drawn from an uniform distribution in the interval $(0,0.25)$. The color bar indicates the local deviations from the corresponding Gaussian copula. Taken from~\cite{Sicking2018}.}
 \label{fig:Copula_Heterogeneous_Sigma_cropped}
\end{figure}

We begin with investigating the heterogeneity of single parameters. Gaussian dynamics with an average asset correlation $c=0.3$ and a homogeneous large negative drift of $\mu = -3\times 10^{-3}\:\text{day}^{-1}$ is considered. Due to the large negative drift we have seen that in the case of an additional homogeneous volatility the resulting dependence structure is a Gaussian copula. A rather simple heterogeneous portfolio is constructed when only the daily volatilities are considered random. For each contract the volatility is drawn from an uniform distribution in the open interval $(0,0.25)$. The resulting average portfolio loss copula is shown in Fig.~\ref{fig:Copula_Heterogeneous_Sigma_cropped}.
We again compare the average copula calculated by the simulation outcomes with the average over the corresponding Gaussian copulas determined by the portfolio loss correlation. Surprisingly, the single parameter heterogeneity is sufficient to cause deviations from the Gaussian copula. The coloring shows deviations of the empirical copula from the Gaussian copula especially in the vicinity of the $(0,0)$ and $(1,1)$ corners. We come to the conclusion that a choice of one or more heterogeneous parameters, \textit{i.e.},~a large variety in different parameters for each portfolio, alters the dependence structure from an ideal Gaussian copula. The more heterogeneous the portfolios become the larger the deviations from the symmetric Gaussian copula.

So far there are two causes for non-Gaussian empirical copulas: the loss of information, induced by the projections of non-defaults onto zero, as well as parameter heterogeneity.

\begin{figure}[htp]
 \centering
 \includegraphics[width=0.5\textwidth]{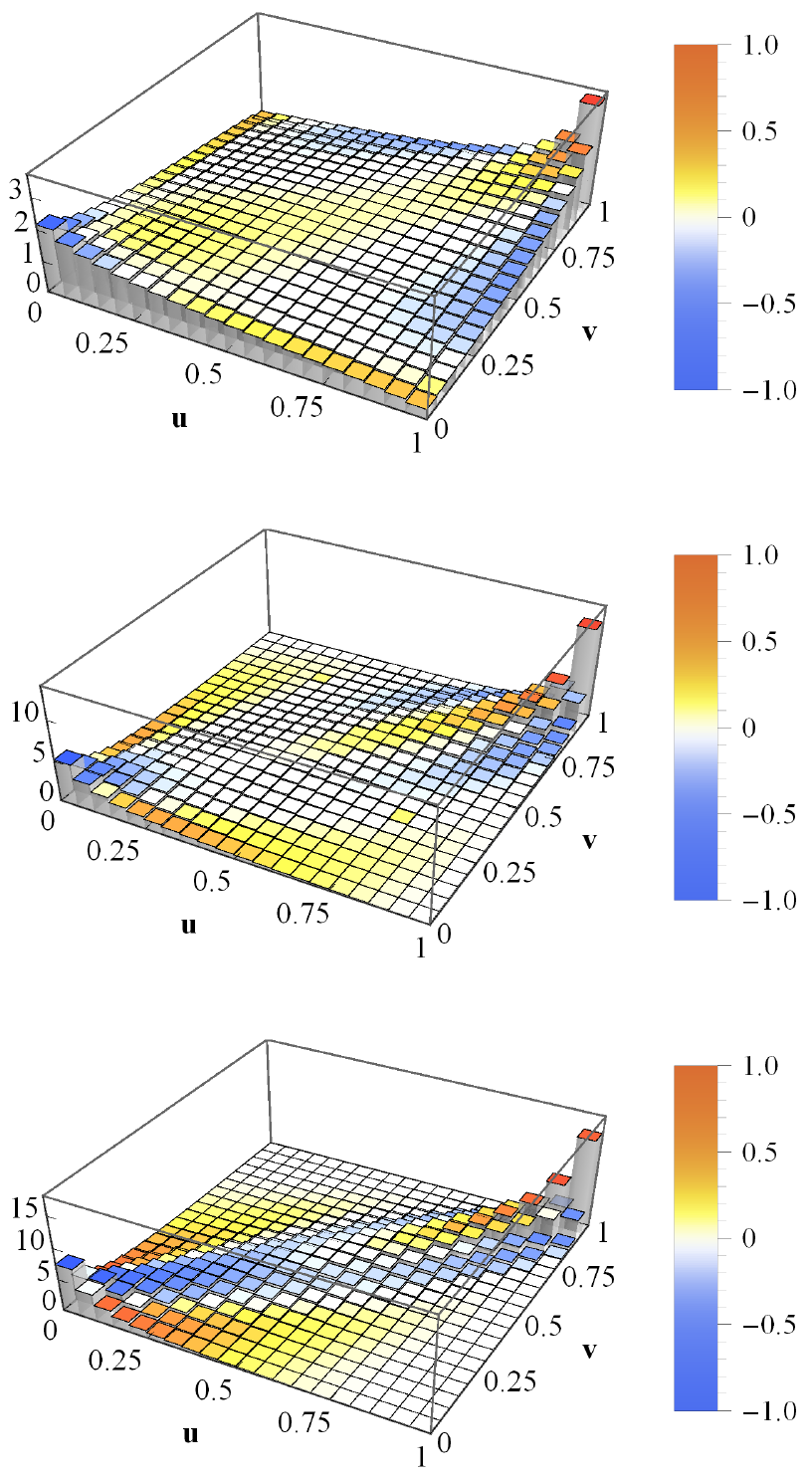}
  \caption{Time averaged loss copula histograms for two empirical copulas of size $K=50$. The asset values are multivariate log-normal ($N\to\infty$). Top: portfolio 1 is always drawn from S\&P 500 and portfolio 2 from Nikkei 225, middle: both portfolios are drawn from S\&P 500, bottom: both portfolios are drawn from Nikkei 225. The color bar indicates the local deviations from the corresponding Gaussian copula. Taken from~\cite{Sicking2018}.}
 \label{fig:Copula_LGD_Copula_Realistic_cropped}
\end{figure}

We now turn to empirical portfolios. Before starting the simulation the empirical parameters have to be defined. The dataset consists of stock return data from 272 companies listed in S\&P~500 index and from 179 companies listed in Nikkei~225 index. It is sampled in a 21-year interval which covers the period 01/1993--04/2014. To set up a realistic, fully homogeneous portfolio, drifts, volatilities and correlations are calculated from this empirical dataset. Moreover in \cite{Schmitt2013} it was shown that annual returns behave normally for empirical asset values. To match these findings the Gaussian dynamics for the stock price returns is applied. To obtain an average empirical portfolio loss copula one first averages over different pairs of portfolios and then averages over randomly chosen annual time intervals taken out of the 21-year period. By averaging over different pairs of portfolios, results which are due to specific features of two particular portfolios are avoided. We consider three different cases which are shown in Fig.~\ref{fig:Copula_LGD_Copula_Realistic_cropped}.
In the first case, which is shown on the top panel, one portfolio is drawn from S\&P~500 and the other is drawn from Nikkei~225. In the second case (middle panel) both portfolios are drawn from S\&P~500 and in the third case (bottom panel) both are drawn from Nikkei~225.

In all three cases we find deviations of the empirical copula from the Gaussian copula. Especially the dependence of the extreme events is much more pronounced than by the prediction of a Gaussian copula. This can be seen in the $(1,1)$ corner, where the colors indicate that the tails are much more narrow and pointed compared to the Gaussian copula. On the other side the tails in the $(0,0)$ corner are flatter compared to a Gaussian copula. The asymmetry regarding the line spanned by $(1,0)$ and $(0,1)$ leads to the conclusion that extreme portfolio losses occur more often simultaneously than in the case of small portfolio losses. Hence, an extreme loss of one portfolio is very likely to also yield an extreme loss of the other portfolio. This dependence is much stronger than predicted by a Gaussian copula. Thus, modeling portfolio loss dependencies by means of Gaussian copulas is deeply flawed and might cause severe underestimations of the actual credit risk. 

Another important aspect of credit risk can be analyzed by considering different portfolio sizes. So far only rather small portfolios of size $K=50$ were chosen. Increasing the size of the portfolios leads to a rise in portfolio loss correlation. This behavior can be explained by the decrease of idiosyncrasy of large portfolios. Moreover it explains why the empirical loss copulas in Fig.~\ref{fig:Copula_LGD_Copula_Realistic_cropped} are almost perfectly symmetric regarding the line spanned by $(0,0)$ and $(1,1)$. Portfolios based on the S\&P~500 dataset with a size of $K=50$ each reveal an significant average loss correlation of $\mathrm{Corr}(L_1,L_2)=0.779$. Even if we decrease the size to $K=14$ companies an average portfolio loss correlation of $\mathrm{Corr}(L_1,L_2)>0.5$ is found. This reveals that high dependencies among banks are not only limited to ``big players'' which hold portfolios of several thousand contracts. Also small institutions show noticeable dependencies even though their portfolios are non-overlapping.

%%%%%%%%%%%%%%%%%%%%%%%%%%%%%%%%%%%%%%%%%%
\section{Discussion}
\label{sec5}

The motivation for the studies which we reviewed here was
twofold. First, the massive perturbations that shook the financial
markets starting with the subprime crisis of 2007-2009 sharpened the
understanding of how crucial the role of credits is for the stability of the
economy as a whole in view of the strong interdependencies. Better
credit risk estimation is urgently called for, particularly for rare
but drastic events, \textit{i.e.},~for the tails of the loss
distributions. Particularly, the often claimed benefit of
diversification has to be critically investigated. Second, the
ubiquitous non-stationarity in financial markets has to be properly
accounted for in the models. The financial crisis illustrates in a
painful way that decisive economic quantities strongly fluctuate in
time, ruling out elegant, but too simplistic equilibrium approaches 
which fail to grasp the empirical situation.

This twofold motivation prompted a random matrix approach to tackle
and eventually solve the Merton model of credit risk for fully
correlated assets. A proper asset value distribution can be
calculated by an ensemble average of random correlation matrices. The
main ingredient is a new interpretation of the Wishart model for
correlation or covariance matrices. While it was originally meant to
describe generic statistical features resulting from
\textit{stationary} time series, \textit{i.e.}, eigenvalue densities
and other quantities for large correlation matrices, the new
interpretation grasps \textit{non-stationary} correlation matrices by
modeling a truly existing, measured set of such correlation matrices
with an ensemble of random correlation matrices. Contrary to the
original interpretation of the Wishart model, ergodicity reasoning is
not applied, and a restriction to large matrices is not needed,
either.

According to the Merton model, stock price data instead of data on asset values
can be used to calibrate the required parameters. This is quite valuable because
empirical data on asset values is not easy to obtain, whereas stock
price data is commonly available. Considering long time horizons, the
sample statistics of returns can be described by a multivariate
mixture of distributions. The resulting distribution is the average of
a multivariate normal distribution over an ensemble of
Wishart-distributed covariance matrices. This random matrix approach takes
the fluctuating asset correlations into account. As a very nice side
effect the random matrix approach reduces the number of parameters which
describe the correlation structure of the financial market onto
two. Both of them can be considered as macroscopic. One parameter is a
mean correlation coefficient of the asset values and the other
parameter describes the strength of the fluctuations around this
average. Furthermore, the random matrix approach yields analytical
tractability which allows to derive an analytical expression for the
loss distribution of a portfolio of credit contracts, taking
fluctuating asset correlations into account. In a quantitative manner
it is shown that in the presence of asset correlations diversification
fails to reduce the risk of large losses. This is substantial
quantitative support and corroboration for qualitative reasoning in
the economic literature. Furthermore, it is demonstrated that the
random matrix approach can describe very different market situations.
For example, in a crisis, the mean correlation coefficient is higher
and the parameter governing the strength of the fluctuations is
smaller than in a quite period, with considerable impact on the loss 
distribution.

In addition Monte-Carlo simulations were run to calculate VaR and
ETL. The results support the approximation of an effective average
correlation matrix if heterogeneous average volatilities are taken
into account. Moreover the simulations show the benefit of the
random matrix approach. If the fluctuations between the asset correlations
are neglected the VaR is underestimated by up to 40\%. This
underestimation could yield dramatic consequences. Therefore the
results strongly support a conservative approach to capital reserve
requirements.

Light is shed on intrinsic instabilities of the financial
sector. Sizable systemic risks are present in the financial
system. These were revealed in the financial crisis 2007--2009. Up to
that point tail-dependencies between credit contracts were
underestimated which emerged as a big problem in credit risk
assessment. This is another motivation for models like ours that take
asset fluctuations into account.

The dependence structure of credit portfolio losses was analyzed
within the framework of the Merton model. Importantly, the two credit
portfolios operate on the same correlated market, no matter if they
belong to a single creditor or bank or to different banks. The
instruments to analyze the joint risk are correlations and
copulas. Correlations break down the dependence structure onto one
parameter and represent only a rough approximation for the coupling of
portfolio losses. In contrast copulas reveal the full dependence
structure.  For two non-overlapping credit portfolios we found
concurrent large portfolio losses to be more likely than concurrent
small ones. This observation is in contrast to a symmetric Gaussian
behavior as described by correlation coefficients. Risk estimation by
solely using standard correlation coefficients yields a clear
underestimation of concurrent large portfolio losses. Hence, from a
systemic viewpoint it is really necessary to incorporate the full
dependence structure of joint risks.

%%%%%%%%%%%%%%%%%%%%%%%%%%%%%%%%%%%%%%%%%%

\vspace{6pt}

%%%%%%%%%%%%%%%%%%%%%%%%%%%%%%%%%%%%%%%%%%
\section{Acknowledgement}
A.M.~acknowledges support from Studienstiftung des deutschen Volkes.

%\reftitle{References}

%%%%%%%%%%%%%%%%%%%%%%%%%%%%%%%%%%%%%%%%%%
\end{document}